\documentclass[11pt]{article}

\usepackage{amsmath,amsthm,amssymb}
\usepackage{accents}
\usepackage[utf8]{inputenc}
\usepackage{mathtools}
\usepackage[colorlinks=false, urlcolor=blue]{hyperref}
\usepackage[round]{natbib}
\usepackage{setspace,rotating}
\usepackage{threeparttable,booktabs}
\usepackage{graphics,graphicx}
\usepackage{chngcntr}

\newtheorem{proposition}{Proposition}

\newcommand{\Var}{\mathrm{Var}}
\newcommand{\Cov}{\mathrm{Cov}}
\newcommand{\plim}{\mathrm{plim}}

\hypersetup{
	colorlinks   = true, 
	urlcolor     = blue, 
	linkcolor    = blue, 
	citecolor   = blue 
}

\title{Bounds for Bias-Adjusted Treatment Effect in Linear Econometric Models}

\author{Deepankar Basu\thanks{Department of Economics, University of Massachusetts Amherst. Email: dbasu@econs.umass.edu. I would like to thank Ina Ganguli and Leila Gautham for comments on an earlier version of this paper and Evan Wasner for excellent research assistance.}}

\begin{document}
	
	\maketitle
	
	\begin{abstract}
		\noindent
		In linear econometric models with proportional selection on unobservables, omitted variable bias in estimated treatment effects are real roots of a cubic equation involving estimated parameters from a short and intermediate regression. The roots of the cubic are functions of $\delta$, the degree of selection on unobservables, and $R_{max}$, the R-squared in a hypothetical long regression that includes the unobservable confounder and all observable controls. In this paper I propose and implement a novel algorithm to compute roots of the cubic equation over relevant regions of the $\delta$-$R_{max}$ plane and use the roots to construct bounding sets for the true treatment effect. The algorithm is based on two well-known mathematical results: (a) the discriminant of the cubic equation can be used to demarcate regions of unique real roots from regions of three real roots, and (b) a small change in the coefficients of a polynomial equation will lead to small change in its roots because the latter are continuous functions of the former. I illustrate my method by applying it to the analysis of maternal behavior on child outcomes. 
	\end{abstract}
	
	
	\section{Introduction}	
	Researchers are often interested in estimating treatment effects in models where there are clear problems of unobserved or unobservable confounders. Such hypothetical `long' regressions cannot be estimated because of unobservability of the confounding regressor. Faced with this problem, researchers often compare the ordinary least square (OLS) estimate of the treatment effect between a `short' and an `intermediate' regression, both of which can be estimated. In the short regression both the observable and unobservable controls are left out; in the intermediate regression only the unobservable control is missing from the model. If the numerical magnitudes of the treatment effect are roughly similar in both the short and intermediate regressions, i.e. the estimate of the treatment effect is `stable', then researchers conclude that the bias from the omitted, unobservable confounder is small. 
	
	In a recent, innovative contribution, \citet{oster-2019} has demonstrated that such `coefficient stability' arguments to deal with possible omitted variable bias is misleading.\footnote{\citet{oster-2019} extends previous work on this issue by \citet{aet-2000,aet-2005}.} In fact, what is needed to draw conclusions about the magnitude of possible bias due to the unobservable confounder is not the raw change in the estimate of the treatment effect, but an R-squared scaled change in the estimate of the treatment effect between the short and intermediate regressions. This becomes clear when we write the expression for the omitted variable bias of the treatment effect in the intermediate regression in terms of the R-squared in the short, intermediate and hypothetical long regressions, and relevant coefficients in the long regression. A little algebraic manipulation generates a cubic equation in the bias (of the treatment effect in the intermediate regression). The coefficients of this cubic equation are functions of estimated, and therefore known, parameters and, in addition, two unknown parameters: $\delta$, the relative degree of selection on unobservables, and $R_{max}$, the R-squared in the hypothetical long regression. 
	
	A cubic equation with real (or complex) coefficients will have either one or three real roots. When the cubic equation has a unique real root, the researcher is able to identify the bias, and use it to compute the bias-adjusted treatment effect, without any ambiguity. When the cubic equation has three real roots, the researcher is confronted with the problem of non-uniqueness. Confronted with this problem, previous researchers like \citet{oster-2019}, \citet{aet-2005} and \citet{aet-2000} have proposed approaches that avoids computing roots of the cubic. \citet{oster-2019} proposes two approaches to deal with the problem of non-uniqueness that does not require the researcher to compute roots of the cubic equation. Unfortunately, these methods suffer from some serious shortcoming (which I point out later in this introductory section and then discuss in greater detail in section~\ref{sec:eqsel}).
	
	I propose an alternative methodology to quantify the bias in treatment effect. At the center stage of my method is an algorithm to compute and select the correct real root of the cubic equation. To understand my proposed algorithm let us return to the cubic equation that is at the heart of the bias calculations. The coefficients of the cubic equation are functions of two unknown parameters: $\delta$, and $R_{max}$. In any given econometric analysis, details of the question under investigation will allow a researcher to choose a plausible \textit{range} (an interval of the real line) over which $\delta$ and $R_{max}$ can vary.  
	
	To understand how a plausible range can be chosen, note that a value of $\delta$ that is higher than $1$ means that the unobservable confounder is relatively more (less) important than the observed controls in explaining variation in the \textit{treatment variable}. On the other hand, a relatively high (low) value of $R_{max}$ means that the unobservable confounder is relatively more (less) important than the observed controls in explaining variation in the \textit{outcome variable}. Based on the details of the outcome, treatment, included control and omitted variables, a researcher will therefore be able to choose the plausible range for $\delta$ and $R_{max}$. Once this is done, we are able to define a bounded box in the ($\delta$-$R_{max}$) plane as the Cartesian product of the bounded intervals over which $\delta$ and $R_{max}$ span. 
	
	Using the magnitude of the discriminant of the cubic equation we can then divide the bounded box into two parts, the first corresponding to a region where the cubic equation will have unique real roots (this is the region where the discriminant is strictly positive) and the second corresponding to a region where the cubic equation will have three real roots (this is the region where the discriminant is nonpositive). Let us call the first part as the $URR$ (unique real root) area and the second part as the $NURR$ (nonunique real root) area. 
	
	As the first, and simplest case, I consider the situation where the bounded box is completely contained in $URR$. I impose a $N \times N$ grid on the bounded box, compute the unique real roots on all grid points of the box and collect the real roots in a vector, $B_U$, (whose length is equal to $N^2$, the number of points of the grid). We can now compute a bounding set, $S_{URR}$, for the `true' treatment effect as the interval formed by the $2.5$-th  and the $97.5$-th quantile of the empirical distribution of $\beta^*=\tilde{\beta}-\nu$, where $\nu$ are elements of the vector $B_U$. This is because the unique real root is the bias of the treatment effect (estimated from the intermediate regression). Hence, the `true' treatment effect is the difference between the treatment effect estimated from the intermediate regression, $\tilde{\beta}$, and the root of the cubic, $\nu$. Thus, the difference between $\tilde{\beta}$ and the vector, $B_U$, gives us a $N^2 \times 1$ vector of `true' treatment effects. And since we have covered the whole bounded box, as long as the assumptions generating the bounded box are correct, the empirical distribution of $\beta^*$, the bias-adjusted treatment effect (BATE), will give us a good \textit{approximation} of the 95\% confidence interval of $\beta$.
	
	As the next, and computationally more difficult, case, I consider the situation where the bounded box is partly contained in the URR and partly in the NURR. In this situation, I first compute all the unique real roots on $M_1^2$ grid points of the box that lies in URR. Then, I extend the analysis to the NURR by taking recourse to a result from complex analysis \citep{alexanderian-2013}: the roots of a polynomial are continuous functions of the coefficients of the polynomial; moreover, real roots of multiplicity one will remain real when the coefficients are perturbed slightly. 
	
	After computing the $M_1^2$ unique real roots on URR, I impose a grid of $M_2^2$ points on NURR. I identify border points of the grid that belong to NURR by choosing all points of NURR that are within a pre-specified `small' distance of any point in URR. In each of these border points of NURR, I compute the three real roots and select the one that is closest in absolute value to the \textit{unique} real root computed at its closest grid point in URR. This selection is theoretically justified by the fact that roots of the cubic equation are continuous functions of $\delta$ and $R_{max}$. Hence, if a grid point in NURR is within a small open ball of a grid point in URR, the unique real root in the latter will be `close' to one of the three real roots in the former. This `closeness' allows me to select \textit{this} root as the correct estimate of bias on the grid point in NURR.
	
	Once the border grid points in NURR are covered, I iterate the process deeper, layer by layer and in `small' steps at a time, into the NURR area until all grid points in NURR are exhausted. At the end of the process, I have therefore generated a $(M_1^2+M_2^2)$-vector of real roots. I use these to compute the bias-adjusted treatment effect, $\beta^*$, exactly as I did in the first case. Using the quantiles of the empirical distribution of $\beta^*$, I can now generate approximate confidence intervals for the `true' $\beta$.
	
	The final case to consider is one when the bounded box is wholly contained in NURR. In this case, the algorithm extends dimensions of the bounded box in the $\delta$ direction in small steps so as to generate a non-empty intersection with the URR. As soon as the algorithm finds a non-empty area of intersection with URR, it then applies the logic of the second case to complete the analysis.
	
	The choice of bounds for $\delta$ and $R_{max}$ define the meaningful area over which roots of the relevant cubic equation is solved. Hence, the results of the analysis that I propose depend crucially on these bounds. Since $\delta$ captures the relative importance of the unobservable confounder compared to the observed controls in explaining the variation in the treatment variable, we can define a low $\delta$ regime as one where $0 < \delta <1$ and a high $\delta$ regime as one where $1 < \delta <\delta_{high}$.\footnote{The choice of $\delta_{high}$ will be decided by the researcher. What we need is a large positive number, significantly higher than unity, which ensures that the box has non-empty intersection with URR.} On the other hand, $R_{max}/\tilde{R}$ captures the relative importance of the unobserved confounder compared to the observed controls in explaining the variation of the outcome variable (recall that $R_{max}$ and $\tilde{R}$ are the R-squared of the intermediate and long regressions). Since the hypothetical regressions has more regressors than the intermediate regression, $\tilde{R} \leq R_{max} \leq 1$. While the lower bound of $R_{max}$ is thereby fixed, the upper bound will need to be chosen by the researcher. A value of $1$ is too restrictive because even in the hypothetical long regression, the regressors cannot be expected to explain all the variation in the outcome variable (perhaps due to measurement error). Hence, we must use some upper bound for $R_{max}$ that is lower than unity. While \citet[section~5.2]{oster-2019} proposes an upper bound of $1.3\tilde{R}$, researchers can also try other numbers that are theoretically or empirically justified.
	
	My proposed method has clear advantages over the two ways to address nonuniqueness that was proposed in \citet{oster-2019}. The first method proposed by \citet{oster-2019} involves computing the bias-adjusted treatment effect under the twin assumptions of $\delta=1$ (equal selection on observables and unobservables) and a sign restriction (which is stated as Assumption~$ 3 $ in the paper). In section~\ref{sec:eqsel}, I show that even on theoretical grounds this does not solve the nonuniqueness problem. I also demonstrate the problem using real data in section~\ref{sec:oster-idset}. The second method proposed by \citet{oster-2019} relies on choosing some value of $R_{max}$, and calculating the magnitude of $\delta$, i.e. degree of selection due to unobservables, that would be consistent with $\beta=0$ (no treatment effect). While this method solves the nonuniqueness issue, it suffers from serious problems of robustness and interpretation, as I discuss in section~\ref{sec:deltastar}.

	In comparison to the first method of \citet{oster-2019}, my method is free of the theoretical problems that I identify in her method. My method offers a transparent, theoretically grounded method of computing bounds for the `true' treatment effect. In comparison to the second method proposed by \citet{oster-2019}, my method is more robust. Instead of choosing a specific value of $R_{max}$, as \citet{oster-2019} does, I compute and then use the treatment bias for all possible values of $R_{max}$ and $\delta$ over a meaningful area. While the methodology of \citet{oster-2019} can be extremely sensitive to the correct choice of $R_{max}$, my method is less so because it relies on computing bias over a whole region. 
		
	After presenting the theoretical results, I use my method on a data set that comes from the Children and Young Adults sample of the NLSY and is used to study the impact of maternal behaviour on child outcomes.\footnote{I would like to thank Emily Oster for making her data set available. I have downloaded the data set from her webpage: \url{https://drive.google.com/file/d/0B1U4uS7GkkxbV0VkZmd0ZVlDVDA/view?usp=sharing}} Using this data set, I highlight, at various points in the paper, the differences in my methodology from the results reported and discussed in \citet[section~4.2]{oster-2019}. The algorithm proposed in this paper has been implemented in an R package called \texttt{bate} (bias adjusted treatment effect) and can be downloaded from \url{https://github.com/dbasu-umass/bate}.
	
	The rest of the paper is organized as follows. In section~\ref{sec:basic}, I discuss the basic set-up; in section~\ref{sec:cubic}, I present my method of analyzing bias and discuss details of the proposed algorithm; in section~\ref{sec:appl}, I illustrate my method, using a data set (NLSY) to study the impact of maternal behaviour on child outcomes; in section~\ref{sec:oster-compare}, I compare my approach with Oster's and point out some problems in the latter; in section~\ref{sec:concl}, I conclude with a summary of my proposed methodology.
	
	\section{Basic Set-Up}\label{sec:basic}
	\subsection{Four Regression Models}
	Consider a hypothetical `long' regression, 
	\begin{equation}
	Y = \beta X + \Psi \omega^0 + W_2 + \varepsilon,
	\end{equation}
	where $Y$ is the scalar outcome variable, $X$ is the scalar treatment variable, $\omega^0$ is a $k$-vector of observable control variables, $W_2$ is the unobserved control variable (understood as an index of any number of unobserved control variables), $\varepsilon$ is a stochastic error term, $\alpha$ is a scalar parameter, $\beta$ is the scalar parameter of interest to the researcher (which captures the treatment effect) and $\Psi$ is a $1 \times k$ vector of parameters. Let us denote the R-squared from this hypothetical long regression as $R_{max}$ and note that $R_{max}$ is unknown because this regression cannot be estimated (because $W_2$ is unobserved). 
	
	The researcher, instead, estimates an `intermediate' regression, by regressing $Y$ on $X$ and $\omega^o$ (where $W_2$ is left out of the regression). Let us denote the estimated coefficient on $X$ and the R-squared in the intermediate regression as $\tilde{\beta}$ as $\tilde{R}$, respectively. For analytical purposed, let us also consider two more regression models: (a) a `short' regression, where $Y$ is regressed on $X$; and (b) an auxiliary regression, where $X$ (treatment variable) is regressed on $\omega^o$ (all the observable control variables). Let us denote the estimated coefficient on $X$ and the R-squared from the short regression as $\mathring{\beta}$ and $\mathring{R}$, respectively; let us denote by $\tau_X$, the variance of the residual from the auxiliary regression. Finally, let $\sigma_X^2$ denote the variance of $X$ (treatment variable), and let $\sigma_Y^2$ denote the variance of $Y$ (outcome variable).

	\subsection{Proportion of Selection}
	Following \citet{oster-2019}, let us define the measure of proportional selection on unobservables as,
	\begin{equation}
	\delta = \frac{\sigma_{2X}/\sigma^2_2}{\sigma_{1X}/\sigma^2_1}
	\end{equation}
	where $\sigma_{1X}=\Cov\left( W_1, X\right) $, $\sigma_{2X}=\Cov\left( W_2, X\right) $, $\sigma^2_1=\Var\left( W_1\right) $, and $\sigma^2_2=\Var(W_2)$, and $W_1=\Psi \omega^0$ (an index of the observable controls). Let us try to understand the meaning of this parameter, $\delta$? 
	
	Consider a linear projection \citep[chapter~2]{wooldridge} of the treatment variables on the index of the observables, i.e.
	\[
	X = \alpha_0 + \alpha_1 W_1 + u_1.
	\] 
	Since $u_1$ is orthogonal to $W_1$ by definition of linear projections, we have
	\begin{equation}
	\alpha_1 = \frac{\sigma_{1X}}{\sigma^2_1}.
	\end{equation}
	Now consider another linear projection of the treatment variables on the index of the unobservables, i.e.
	\[
	X = \delta_0 + \delta_1 W_2 + u_2
	\] 
	and note, once again using the property of linear projections, that
	\begin{equation}
	\delta_1 = \frac{\sigma_{2X}}{\sigma^2_2}.
	\end{equation}
	Now we see clearly that the measure of proportional selection is just the ratio of the two coefficients from the two linear projections, i.e.
	\begin{equation}\label{prop-sel}
	\delta = \frac{\delta_1}{\alpha_1}.
	\end{equation}
	We will return to this expression when we try to look critically at the use of $\delta=1$ as a lower bound.

	 \subsection{Cubic Equation in Bias}
	 Let $\nu$ denote the asymptotic bias in the treatment effect estimated from the intermediate regression, i.e.
	 \begin{equation}
	 \nu = \plim \tilde{\beta} - \beta,
	 \end{equation}
	 where $\plim \tilde{\beta}$ denotes the probability limit of $\tilde{\beta}$ as the sample size increases without bound.\footnote{Note that all estimators in this paper are functions of the sample size, $N$. We suppress this dependence for notational simplicity.} For $j = 1,2, \ldots, J$, let $\omega_{jj}=\Var(\omega^o_j)$ denote the variance of the $j$-th observed control variable, for $j \neq k$, let $\omega_{jk}=\Cov(\omega^o_j,\omega^o_k)$ denote the covariance between the $j$-th and $k$-th observed control variables, and for $j = 1,2, \ldots, J$, let $\sigma_{2j}=\Cov(W_2,\omega^o_j)$ denote the covariance between $W_2$ (the index of unobserved confounders) and the $j$-th observed control variable.
	 \begin{proposition}\label{prop:3eqns}
	 	If, for $j,k = 1,2, \ldots, J$, $j \neq k$, 
	 	\begin{equation}\label{cond1}
	 	\omega_{jk}=0,
	 	\end{equation}
	 	and for $j = 1,2, \ldots, J$, 
	 	\begin{equation}\label{cond2}
	 	\sigma_{2j}=0,
	 	\end{equation}
	 	then we have,
	 	\begin{equation}\label{eq:sol1}
	 	\left( \mathring{\beta} - \tilde{\beta}\right) \overset{p}{\to} \frac{\sigma_{1X}}{\sigma_X^2} - \nu \left( \frac{\sigma_X^2 - \tau_X}{\sigma_X^2}\right), 
	 	\end{equation}
	 	
	 	\begin{equation}\label{eq:sol2}
	 	\left( \tilde{R} - \mathring{R}\right) \sigma_y^2 \overset{p}{\to} \sigma_1^2 + \tau_X \nu^2 - \frac{1}{\sigma_X^2}\left( \sigma_{1X} + \nu \tau_X\right)^2,  
	 	\end{equation}
	 	
	 	\begin{equation}\label{eq:sol3}
	 	\left(R_{max} - \tilde{R}\right) \sigma_y^2 \overset{p}{\to} \nu \left( \frac{\sigma_1^2 \tau_X}{\delta \sigma_{1X}} - \nu \tau_X\right)   
	 	\end{equation}
	 	where $\overset{p}{\to}$ refers to convergence in probability as the sample size increases without bound.
	 \end{proposition} 
	 
	 The details of the proof can be found in Appendix~\ref{cubic-derive}. The usefulness of the above result is that it leads to a cubic equation in the bias. To see this, note that the equations in (\ref{eq:sol1}), (\ref{eq:sol2}) and (\ref{eq:sol3}) constitute a system of $ 3 $ equations in $ 3 $ unknowns: $\sigma^2_1$, the variance of $ W_1 $; $\sigma_{1X}$, the covariance of $ W_1 $ and $ X $ (treatment variable); and $\nu$ (the bias of the treatment effect in the intermediate regression). Algebraic manipulation can reduce the three equations into a single cubic equation in $\nu$ given by,
	 \begin{equation}\label{eq:cubic}
	 a \nu^3 + b \nu^2 + c \nu + d = 0,
	 \end{equation}
	 where,
	 \begin{align}
	 a & = \left( \delta - 1 \right) \left( \tau_X \sigma^2_X - \tau^2_X \right) \neq 0, \label{eq:a} \\
	 b & = \tau_X \left( \mathring{\beta} - \tilde{\beta}\right) \sigma^2_X \left( \delta - 2 \right),  \label{eq:b} \\
	 c & = \delta \left( R_{max} - \tilde{R} \right) \sigma^2_Y \left( \sigma^2_X - \tau_X \right)  - \left( \tilde{R} - \mathring{R} \right) \sigma^2_Y \tau_X \nonumber \\
	 & \qquad - \sigma^2_X \tau_X \left( \mathring{\beta} - \tilde{\beta}\right)^2 \label{eq:c}, \\
	 d & = \delta \left( R_{max} - \tilde{R} \right) \sigma^2_Y \left( \mathring{\beta} - \tilde{\beta}\right) \sigma^2_X. \label{eq:d}  
	 \end{align}
	 
	 This gives us the crucial result about the roots of the cubic equation in (\ref{eq:cubic}).
	 
	 \begin{proposition}\label{prop:consistency}
	 	Consider the cubic equation given in (\ref{eq:cubic}) with coefficients given in (\ref{eq:a}), (\ref{eq:b}), (\ref{eq:c}), and (\ref{eq:d}). 
	 	\begin{enumerate}
	 		\item When the cubic equation has one (unique) real root, denote it by $\nu_{U}$ and let $\beta^*_U=\tilde{\beta}-\nu_U$. Then
	 		\[
	 		\beta^*_U \xrightarrow{p} \beta.
	 		\]
	 		
	 		\item When the cubic equation has three (non-unique) real roots, denote them by $\nu_{NU,1},\nu_{NU,2},\nu_{NU,3}$ and for $i=1, 2, 3$, let $\beta^*_{NU,i}=\tilde{\beta}-\nu_{NU,i}$. Then, for $i=1$ or $i=2$ or $i=3$,
	 		\[
	 		\beta^*_{NU,i} \xrightarrow{p} \beta.
	 		\]
	 	\end{enumerate}
	 \end{proposition}
	 
	 The proof follows immediately from the fact that $\nu$ is \textit{defined to be} the asymptotic bias in the treatment effect estimated from the intermediate regression.\footnote{This result is given in \citet[Proposition~2]{oster-2019}.} For us, it is more important to pay attention to the two maintained assumptions of the whole analysis stated explicitly in Proposition~\ref{prop:3eqns}: (a) the observed controls are pairwise uncorrelated, i.e. for $j \neq k$, $\omega_{jk}=0$, and (b) the unobserved confounder is uncorrelated with each of the observed controls, i.e. $\sigma_{2j}=0$. Both these are strong assumptions and in working out the proof in Appendix~\ref{cubic-derive}, I point out exactly where they are needed. While \citet[pp.~192]{oster-2019} claims that these assumptions do not imply any loss of generality, the derivation in Appendix~\ref{cubic-derive} shows that that is not true. One way to see this is to note that the estimator for the treatment effect, $\beta^*$, is a function of the root of cubic equation in (\ref{eq:cubic}). The cubic equation arises from  (\ref{eq:sol1}), (\ref{eq:sol2}) and (\ref{eq:sol3}), and these three equations, in turn, cannot be derived without the two orthogonality assumptions stated in Proposition~\ref{prop:3eqns}. Hence, the estimator relies crucially on the two orthogonality assumptions, contrary to the argument in \citet[Appendix~A.1]{oster-2019}. Having noted these caveats, let me now turn to the main part of this paper, which is a discussion of a novel algorithm to compute omitted variable bias and bias-adjusted treatment effects (BATE).
	

	\section{Bounds for the Treatment Effect}\label{sec:cubic}
	\subsection{Real Root as Bias and Overall Strategy}
	Finding the real roots of the cubic equation in (\ref{eq:cubic}) is the key to constructing proper bounds for the `true' treatment effect. This follows from Proposition~\ref{prop:consistency}. To do so we note that the coefficients of the cubic equation are composed of all known quantities other than the following two: $R_{max}$ (the R-squared in the hypothetical long regression), and $\delta$ (the measure of proportional selection on unobservables). My strategy consists of the following steps. 
	
	First, I select a bounded box of the $\left( \delta, R_{max}\right) $ plane that is relevant for the econometric analysis in question by choosing lower and upper bounds for $\delta$ and $R_{max}$, i.e. we choose $\delta_{low},\delta_{high}$, and $R_{low},R_{high}$, such that $\delta_{low} < \delta < \delta_{high}$ and $R_{low} < R_{max} < R_{high}$. This defines a bounded box, $B$. Second, I demarcate the portion of the bounded box where the cubic (\ref{eq:cubic}) is guaranteed to have a unique real root (URR) from the portion where it has three real roots (NURR). Third, I impose a sufficiently granular grid of $N^2$ points on B and compute all real roots on the grid points in URR. Fourth, I use continuity of the roots of a polynomial equation with respect to its coefficients to select roots from the grid points in NURR by starting with the roots on the boundary of URR and NURR, and then covering all the grid points on NURR in pre-specified small steps. After I have covered all the $N^2$ points of the grid, I will have an empirical distribution of the omitted variable bias (the selected roots of the cubic equation), and an empirical distribution of the bias-adjusted treatment effect (treatment effect from intermediate regression minus the selected root of the cubic equation).
	
	I will now discuss the details of an algorithm that will implement these ideas.

	\subsection{Algorithm}\label{sec:cubic-algo}
	The algorithm relies on two results, the first relating to the roots of a cubic equation and the second regarding continuity of the roots of any polynomial equation with respect to its coefficients. 
	
	\begin{proposition}\label{prop:cubic-roots}
		Consider the cubic equation in (\ref{eq:cubic}) and let $p=(3ac-b^2)/3a^2$ and $q=(27a^2d + 2b^3 - 9abc)/27a^3$. 
		
		\begin{enumerate}
			\item If $ 27q^2 + 4p^3>0$, then the cubic equation has a unique real root. Let us call the region of the $(\delta, R_{max})$ plane over which this condition is satisfied as URR, the unique real root region. 
			
			\item If $ 27q^2 + 4p^3 \leq 0$, then the cubic equation has three real roots. Let us call the region of the $(\delta, R_{max})$ plane over which this condition is satisfied as NURR, the nonunique real root region. 
		\end{enumerate}
		
	\end{proposition}
	\begin{proof}
		This is a well-known result. See for instance, \citet[Appendix~1]{hellesland-etal-2013} and Appendix~\ref{cubic-solve} for details.
	\end{proof}
	
	The basic idea driving the algorithm is to see how the bounded box, $B$, formed by the choice of the range of values for $\delta$ and $R_{max}$, intersects with the URR and NURR regions, and then to solve the cubic equation on relevant grids covering $B$, starting with the part where $B$ intersects with URR and then extending to NURR using continuity. The last and crucial step of the algorithm, therefore, relies on the following well-known result from complex analysis \citep{alexanderian-2013}: the roots of any polynomial equation are continuous functions of the coefficients of the polynomial, and real roots, upon small changes in the coefficients, remain real. 
	
	\begin{proposition}\label{prop:cont-cubic-roots}
		Consider the cubic equation in (\ref{eq:cubic}), and assume that $b^2 \neq 3ac$. The roots of the cubic equation are continuous functions of $\delta$ and $R_{max}$. Moreover, if $\delta$ and $R_{max}$ are changed by sufficiently small magnitudes, the real roots of multiplicity one remain real.
	\end{proposition}
	\begin{proof}
		Note that the coefficients of cubic equation in (\ref{eq:cubic}) are polynomials in $\delta$ and $R_{max}$. Hence, the coefficients are continuous functions of $\delta$ and $R_{max}$ (because polynomials are everywhere continuous functions). Now, we use the results that the roots of any polynomial are continuous functions of the coefficients of the polynomial \citep{alexanderian-2013}. This implies that the roots are compositions of continuous functions. Hence, they are continuous functions of $\delta$ and $R_{max}$. 
		
		The condition, $b^2 \neq 3ac$, ensures that the real roots have multiplicity of one. To see this, note that the critical points of the cubic polynomial are given by the values of $\nu$ where the first derivative, $3ax^2+2bx+c$, is zero. These are given by
		\[
		\nu_c = \frac{-b \pm \sqrt{b^2 - 3ac}}{3a}.
		\]
		The point of inflection of the cubic is given by the values of $\nu$ where the second derivative, $6ax+2b$, is zero. Hence, it is given by
		\[
		\nu_i = -\frac{b}{3a}.
		\]
		The real root of the cubic equation has multiplicity of $3$ or $1$. A real root has multiplicity of $3$ if and only if $\nu_c=\nu_i$, if and only if $b^2=3ac$. Hence, a necessary and sufficient condition for real roots to have multiplicity of $1$ is that $b^2 \neq 3ac$. This implies, using Theorem 3.5 in \citet{alexanderian-2013}, that small perturbations of $\delta$ and $R_{max}$ will produce \textit{real} roots that are close to the original \textit{real} roots. 
	\end{proof}

	The continuity result is crucial because it allows me to sequentially solve the cubic over grids imposed on $ B $. We start by solving for the cubic equation over grid points in URR (where each point gives a unique real root) and then incrementally move to cover the grid points on the NURR (by selecting among the three real roots that one whose difference in absolute value is least with respect to the unique real root computed on the closest grid point in URR). Since the roots of the cubic equation are continuous in $\delta$ and $R_{max}$, if we are at a point in the NURR that is within a ``small'' distance from a grid point in URR, then the real root at the former will be ``close'' to the unique real root at the latter. This is what allows me to choose one of the three real roots at the former point.
	
	\subsubsection{Case 1}
	In the first case, the bounded box, $B$, is wholly contained in URR. I solve the cubic on a sufficiently granular grid that covers the bounded box. For each point on the grid, using proposition~\ref{prop:cubic-roots}, we can find a unique real root, $\nu$, and use it to construct $\beta^* = \tilde{\beta}-\nu$. If the total number of points on the grid is $M$, we thereby generate a $M$-vector of $\beta^*$ values. The empirical distribution of this $\beta^*$ gives an approximation of the distribution of $\beta$ (the true treatment effect) for the case when the unknown parameters, $\delta$ and $R_{max}$, can range over the chosen bounded box, $B$. Hence, the empirical distribution of $\beta^*$ allows us to construct approximate confidence intervals for $\beta$. The accuracy of the approximation will increase with the number of points in the grid covering the bounded box.
	
	\subsubsection{Case 2}
\begin{figure}
	\centering
	\includegraphics[scale=0.95]{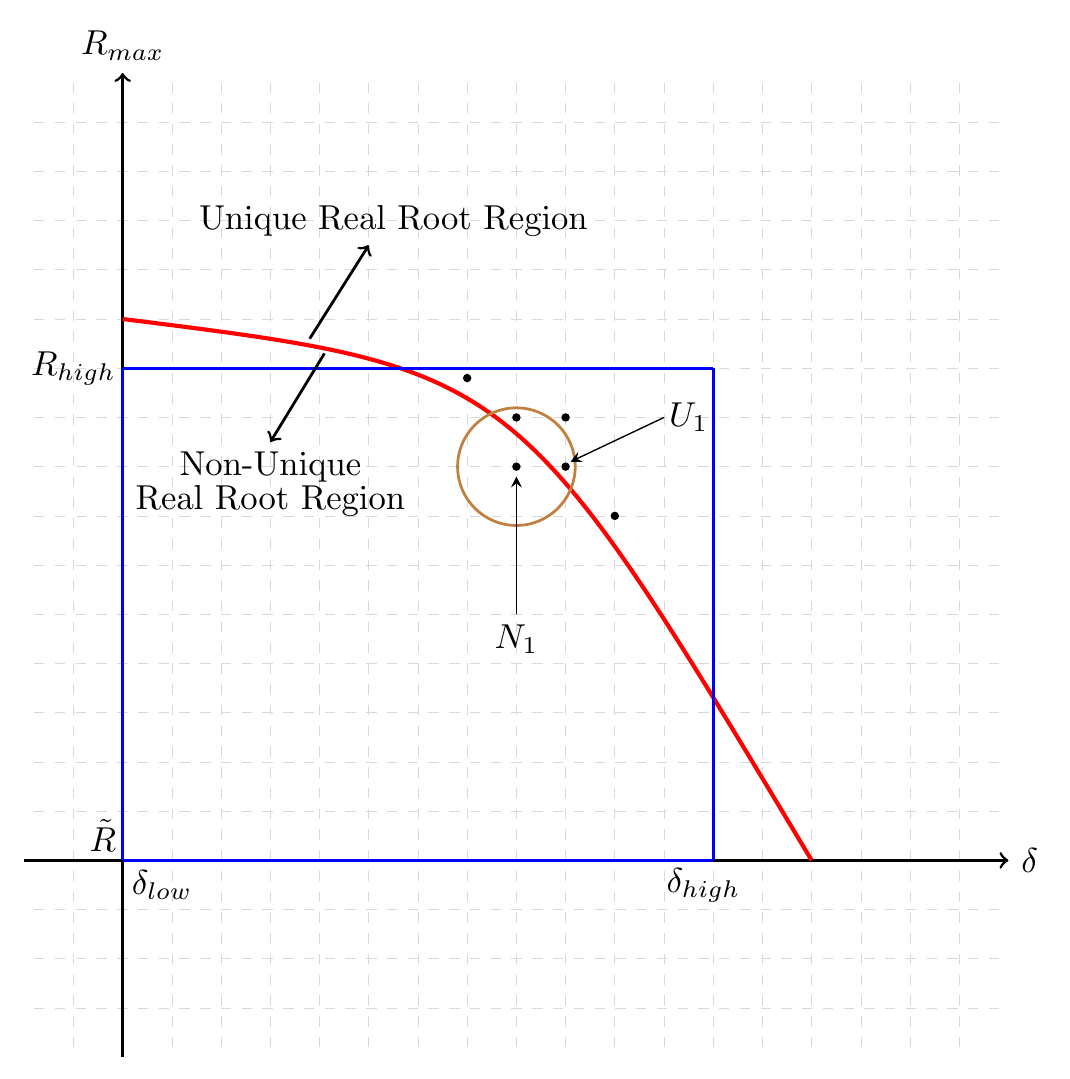}
	\caption{\textit{The figure shows the demarcation of the bounded box (in blue) into the URR (unique real root) and NURR (nonunique real roots) regions. The algorithm selects the real root at the grid point $N_1$, a point in the NURR region, that is closest in absolute value to the unique real root at $U_1$, one of the `close' grid points in the URR region.Continuity of the roots of any polynomial on its coefficients justify this selection. For a discussion, see section~\ref{sec:deltastar}. }}
	\label{fig:algo}
\end{figure}

	In the second case, the bounded box, $B$, is partly contained in URR and partly contained in NURR. This situation is depicted in Figure~\ref{fig:algo}, where $B$ is by the (blue) box contained in the Cartesian product of $[\delta_{low},\delta_{high}]$ and $[\tilde{R}, R_{high}]$. The (red) curve demarcates the plane into URR and NURR: the region above the curve is  URR and the region below is NURR.
	
	\begin{center}
		[Figure~\ref{fig:algo} about here]
	\end{center}
	
	Let $S_1 = B \cap URR$, and $S_2 = B \cap NURR$. For $S_1$, we use the same method as in case 1. This generates, for instance, a $M_1^2$-vector of $\beta^*$ values. The real challenge is to select the `correct' real root for grid points in $S_2$ because each point in $S_2$ generates three real roots. To accomplish this task, we do the following:
	\begin{itemize}
		\item We impose a granular grid on $S_2$ of $M_2^2$ points.
		
		\item We identify points of $S_2$ that are within a pre-specified `small' distance of points in $S_1$, $ e $. What is `small' is determined by the choice of $e$ (which, in turn, determines $M_2$). As $e$ decreases, the distance separating the points on the two sides of the URR/NURR boundary falls. As the distance falls, the computational burden of the method increases, primarily because the number of points in the grid in NURR that has to be compared with points in URR increases and, secondarily because, the cubic equation has to be solved at a larger number of points. This is a trade-off that is inherent to this grid search methodology. 
		
		\item We call the points of $S_2$ identified in the previous step as the set of `closest' points to $S_1$ and denote this set as $S_{21}$. For every point in $S_{21}$, we compute the three real roots of the cubic equation in (\ref{eq:cubic}). We select the real root that is closest, in absolute value, to the unique real root found for the corresponding closest point of $S_1$. This is the `correct' root and is justified by proposition~\ref{prop:cont-cubic-roots}. Figure~\ref{fig:algo} depicts the selection of the `correct' root at a grid point $N_1$, a point in the $NURR$ region, using one of the closest points in the URR region, $U_1$.
		
		\item Next, we identify points of $S_2$ that are within a pre-specified small distance of the set $S_{21}$. We call these the `closest' points to $S_{21}$ and denote this set as $S_{22}$, For every point in $S_{22}$, we compute the three real roots of the cubic equation in (\ref{eq:cubic}). We select the real root that is closest, in absolute value, to the real root that was selected (in the previous step) for the corresponding closest point of $S_{21}$. Once again, this is justified by proposition~\ref{prop:cont-cubic-roots}.
		
		\item We continue this process until we have exhausted all the points in $S_2$. This generates, for instance, a $M_2$-vector of $\beta^*$ values.
		
		\item We combine the $M_1$ and $M_2$ vectors of $\beta^*$ values. We thereby generate a $M$-vector of $\beta^*$ values, where $M=M_1+M_2$,  Now, following the same logic as in case~1, we can generate an empirical distribution of $\beta^*$ and use it to construct approximate confidence intervals for $\beta$.
	\end{itemize}
	
	\subsubsection{Case 3}
	In the third, and final, case, the bounded box, $B$, is wholly contained in NURR. In this case, we extend the dimension of the bounded box, $B$, to the extent that is necessary to generate a non-empty intersection with URR. Once we have generated such a bounded box, we are back to case~2. Hence, we now use the method outlined for case~2 to compute the empirical distribution of the bias and $\beta^*$.

	\section{An Application}\label{sec:appl}
	
	In this section, I report results of applying my method to the analysis of maternal behavior on child outcomes that was discussed in \citet[Section 4.2]{oster-2019}. The substantive issue under investigation in this application is the impact of maternal behaviour on child outcomes. In particular, two child outcomes are studied: a child's standardized IQ score and a child's birth weight. In the study of child IQ, three treatment variables are used in turn: months of breastfeeding, any drinking of alcohol in pregnancy, and an indicator for being low birthweight and preterm. In studying child birthweight, two treatment variables are used, one by one: maternal smoking during pregnancy, and maternal drinking during pregnancy. The following control variables are used for both studies: child race, maternal age, maternal education, maternal income, maternal marital status. The question of interest is whether the treatment variables, each on their own, have any causal impact on the outcome variables.
	
	\subsection{Analysis of Bias and Bounding Sets}
	In Table~\ref{table:maternal-coeff}, I present the estimates of the treatment effect from the short and intermediate regressions. These results replicate the corresponding results in Table~3 in \citet[Column~1, 2]{oster-2019}. For instance, if we look at the first row of Table~\ref{table:maternal-coeff}, we see that the effect of (months of) breastfeeding on child IQ is $ 0.044 $ (column~1) in the short regression and $ 0.017 $ (column~4) in the intermediate regression. Thus, breastfeeding has a positive impact on a child's IQ score. Moving from the short to the intermediate regression, the R-squared increases from $ 0.045 $ (column~3) to $ 0.256 $ (column~6). We can read all the other numbers in columns~1 through ~6 in a similar manner. Since these models are likely to have omitted variables, we would like to quantify the effect of the omitted variable bias. 
	
	\begin{center}
		[Table~\ref{table:maternal-coeff} about here]
	\end{center}

	I begin the analysis of bias by constructing two bounded boxes on the ($\delta, R_{max}$) plane. Box~1 is given by the Cartesian product of ($0.01<\delta<0.99$) and ($\tilde{R}<R_{max}<0.61$), and Box~2 is defined by the Cartesian product of ($1.01<\delta<3.99$) and ($\tilde{R}<R_{max}<0.61$). I use \textit{two} bounded boxes so that I can compare results between a case when the relative selection on unobservables, $\delta$, is lower than $1$ to a case when it is larger than $1$. The lower limit of $R_{max}$ comes from the result that $\tilde{R} \leq R_{max}$ because the hypothetical long regression has at least one more regressor than the intermediate regression. The upper limit of $R_{max}=0.61$ (first three regression models) and $R_{max}=0.53$ (last two regression models) is taken to tally with the same assumption in \citet[Table~3]{oster-2019}.

	\begin{center}
	[Table~\ref{table:bset-maternal} about here]
	\end{center}
	
	On each bounded box, I use proposition~\ref{prop:cubic-roots} to identify the $ URR $ and $NURR$ areas. To construct the grid, I use a step size of $0.01$ and then use the algorithm of section~\ref{sec:cubic-algo} to construct empirical distributions of the omitted variable bias and the bias-adjusted treatment effect (BATE). I summarize the results of this bounding analysis in Table~\ref{table:bset-maternal}. Region plots showing the demarcation of the bounded boxes into URR and NURR, and contour plots of the estimated bias on the bounded boxes are presented in Figure~\ref{fig:mat11} through Figure~\ref{fig:mat52} in the appendix.
	
	\subsubsection{Effect of Breastfeeding on Child IQ}
	The first four rows of Table~\ref{table:bset-maternal} refer to the first row in Table~\ref{table:maternal-coeff} and also to row 1 in \citet[Table~3]{oster-2019}. In this case, the outcome variable is a child's IQ score and the treatment variable is months of breastfeeding. The first four rows of Table~\ref{table:bset-maternal} report quantiles of the empirical distribution of bias and BATE computed over the two bounded boxes displayed in Figure~\ref{fig:mat11} and ~\ref{fig:mat12} in the appendix. From the second row of Table~\ref{table:bset-maternal}, we can see that an approximate 95\% confidence interval for the treatment effect would be $(-0.022, 0.017)$ if we chose to use the first bounded box (Box~1) as the relevant region over which to conduct the bounding exercise. On the other hand, if we chose to use the second bounded box (Box~2) as the relevant region over which $\delta$ and $R_{max}$ can vary, then the approximate 95\% confidence interval for the treatment effect is given by $(-0.116, 0.017)$. In both cases, the substantive conclusion would be that the treatment effect can be completely nullified once the effect of omitted variables are taken into account.

	\subsubsection{Effect of Drinking during Pregnancy on Child IQ}
	The second block of four rows of Table~\ref{table:bset-maternal} refer to the second row in Table~\ref{table:maternal-coeff} and the same row in \citet[Table~3]{oster-2019}. In this case, the outcome variable is the same as in the previous analysis: a child's IQ score. The treatment variable is whether the mother reported drinking alcohol during pregnancy. From the sixth row of Table~\ref{table:bset-maternal}, we can see that an approximate 95\% confidence interval for the treatment effect would be $(-0.104, 0.050)$ if we chose to use the first bounded box (Box~1) as the relevant region over which to conduct the bounding exercise. On the other hand, if we chose to use the second bounded box (Box~2) as the relevant region over which $\delta$ and $R_{max}$ can vary, then the approximate 95\% confidence interval for the treatment effect is given by $(-1.107, 0.050)$. In both cases, once again, the substantive conclusion would be that the treatment effect is likely to be completely nullified once the effect of omitted variables are taken into account.

	\subsubsection{Effect of LBW + Preterm on Child IQ}
	The third block of four rows of Table~\ref{table:bset-maternal} refer to the third row in Table~\ref{table:maternal-coeff} and the same row in \citet[Table~3]{oster-2019}. In this case, the outcome variable is the same as in the previous analysis: a child's IQ score. The treatment variable is whether the mother had low birth weight and was prematurely born. From the tenth row of Table~\ref{table:bset-maternal}, we can see that an approximate 95\% confidence interval for the treatment effect would be $(-0.125, -0.054)$ if we chose to use the first bounded box (Box~1) as the relevant region over which to conduct the bounding exercise. On the other hand, if we chose to use the second bounded box (Box~2) as the relevant region over which $\delta$ and $R_{max}$ can vary, then the approximate 95\% confidence interval for the treatment effect is given by $(-0.125, 0.175)$. The substantive conclusion now changes depending on which box the researcher uses. If Box~1 is the relevant region to conduct the bounding exercise, then the treatment effect will remain negative and significantly different from zero even after we have accounted for omitted variable bias. On the other hand, if Box~2 is the relevant region to use, we cannot rule out the fact that the treatment effect might be completely nullified once the effect of omitted variables are taken into account.
	
	\subsubsection{Effect of Smoking during Pregnancy on Child Birth Weight}
	The fourth block of four rows of Table~\ref{table:bset-maternal} refer to the fourth row in Table~\ref{table:maternal-coeff} and the same row in \citet[Table~3]{oster-2019}. In this case, the outcome variable is a child's birth weight. The treatment variable is whether the mother reported smoking during pregnancy. From the fourteenth row of Table~\ref{table:bset-maternal}, we can see that an approximate 95\% confidence interval for the treatment effect would be $(-172.261, -21.121)$ if we chose to use the first bounded box (Box~1) as the relevant region over which to conduct the bounding exercise. On the other hand, if we chose to use the second bounded box (Box~2) as the relevant region over which $\delta$ and $R_{max}$ can vary, then the approximate 95\% confidence interval for the treatment effect is given by $(-2070.193, 117.522)$. In first case, when $\delta$ is restricted to lie between $0$ and $1$, the treatment effect is likely to remain intact, i.e. different from zero, even when the effect of omitted variables are taken into account. On the other hand, if $\delta$ is allowed to be larger than $1$, then the approximate 95\% confidence interval of the empirical distribution of the treatment effect contains zero. Hence, we cannot rule out the fact that allowing for the effect of the omitted variables will wipe out the treatment effect.

	\subsubsection{Effect of Drinking during Pregnancy on Child Birth Weight}
	The fifth block of four rows of Table~\ref{table:bset-maternal} refer to the fifth row in \citet[Table~3]{oster-2019}. In this case, the outcome variable is a child's birth weight. The treatment variable is whether the mother reported smoking during pregnancy. From the eighteenth row of Table~\ref{table:bset-maternal}, we can see that an approximate 95\% confidence interval for the treatment effect would be $(-14.149, 2.895)$ if we chose to use the first bounded box (Box~1) as the relevant region over which to conduct the bounding exercise. On the other hand, if we chose to use the second bounded box (Box~2) as the relevant region over which $\delta$ and $R_{max}$ can vary, then the approximate 95\% confidence interval for the treatment effect is given by $(-13.149, 138.120)$ in the twentieth row of Table~\ref{table:bset-maternal}. Here we have another example where the substantive conclusion \textit{does not} depend on which box is chosen as the correct one. Irrespective of whether we choose Box~1 or Box~2, the conclusion would be that the treatment effect is likely to be completely nullified once the effect of omitted variables are taken into account. 
	
	\subsection{Step size of grid?}
	The choice of the step size of the grid over which the bias is computed needs to balance an important trade off. On the one hand, the smaller the size of the steps used in constructing the grid, the better the approximation of the bounding set to the the true confidence interval. On the other hand, the smaller the size of the steps, the larger than number of grid points. Hence, the more computationally intensive the implementation of the algorithm. To assess the step size on this trade off, in Table~\ref{table:bdset-e}, I report the confidence interval for the treatment effect for different step sizes. For this exercise, I use the model where the outcome variable is the child's IQ score and the treatment variable is the months of breast feeding (this case is reported in row~1, Table~\ref{table:maternal-coeff}).
	
	\begin{center}
		[Table~\ref{table:bdset-e} about here]
	\end{center}
	
	In Table~\ref{table:bdset-e}, I report the quantiles of the empirical distribution of $\beta^*$ (the bias-adjusted treatment effect) for step sizes of $e=1/25$, $e=1/50$, $e=1/100$, $e=1/250$, and $e=1/500$. From the results in the table, we see that the quantiles of the empirical distribution of $\beta^*$ remains largely unchanged for step sizes lower than $e=1/50$. On the other hand, the computing time increases rapidly as the step size is reduced. In terms of the computation-accuracy trade off, a choice of $e=1/100$ seems to good because it gives us a fairly accurate confidence interval without consuming too much computing power. This is the rationale behind by choice of $e=1/100$ for the analyses reported in Table~\ref{table:bset-maternal}.  
	
	\section{Comparison with Oster's Methodology}\label{sec:oster-compare}
	\citet{oster-2019} proposed two methods for dealing with the problem of non-uniqueness. The first involves constructing identified sets under the assumption of equal selection, i.e. $\delta=1$, and the second involves computing the value of $\delta$ that is necessary to force the treatment effect to become zero. I would now like to point towards some problems in both these methods.
	
	\subsection{Identified Sets Under Equal Selection}\label{sec:eqsel}
	
	\subsubsection{Bias Under Equal Selection}
	The first method proposed by \citet{oster-2019} is to compute `identified' sets under the assumption of equal selection, i.e. $\delta=1$. To compute these `identified' sets, one has to first solve for the bias under equal selection. If we impose the restriction that $\delta=1$ on the coefficients of the cubic in (\ref{eq:cubic}) we get, 
	\begin{align*}
	a & = 0\\
	b & = -\tau_X \left( \mathring{\beta} - \tilde{\beta}\right) \sigma^2_X   \\ 
	c & = \left( R_{max} - \tilde{R} \right) \sigma^2_Y \left( \sigma^2_X - \tau_X \right)  - \left( \tilde{R} - \mathring{R} \right) \sigma^2_Y \tau_X - \sigma^2_X \tau_X \left( \mathring{\beta} - \tilde{\beta}\right)^2  \\ 
	d & = \left( R_{max} - \tilde{R} \right) \sigma^2_Y \left( \mathring{\beta} - \tilde{\beta}\right) \sigma^2_X,  
	\end{align*}	
	which converts the cubic in (\ref{eq:cubic}) to a quadratic equation in $\nu$,
	\begin{equation}\label{eq:quad}
	b_1 \nu^2 + c_1 \nu + d_1 = 0,
	\end{equation}
	where the coefficients of this quadratic are given by, 
	\begin{align}
	b_1 & = -\tau_X \left( \mathring{\beta} - \tilde{\beta}\right) \sigma^2_X   \label{eq:bnew} \\ 
	c_1 & = \left( R_{max} - \tilde{R} \right) \sigma^2_Y \left( \sigma^2_X - \tau_X \right)  - \left( \tilde{R} - \mathring{R} \right) \sigma^2_Y \tau_X - \sigma^2_X \tau_X \left( \mathring{\beta} - \tilde{\beta}\right)^2 \label{eq:cnew} \\ 
	d_1 & = \left( R_{max} - \tilde{R} \right) \sigma^2_Y \left( \mathring{\beta} - \tilde{\beta}\right) \sigma^2_X. \label{eq:dnew} 
	\end{align}
	The solutions of the quadratic in (\ref{eq:quad}) are given by
	\[
	\nu = \frac{-c_1 \pm \sqrt{c_1^2 - 4d_1b_1}}{2b_1},
	\]
	which are noted in Corollary~1 in \citet[pp.~193]{oster-2019}. 
	
	\begin{proposition}\label{prop1}
		The quadratic equation in (\ref{eq:quad}) either has a unique real root or two distinct real roots. It does not have any complex roots.
	\end{proposition}
	\begin{proof}
		The proof follows by noting that the discriminant of this quadratic equation is non-negative, i.e. $c_1^2-4d_1b_1 \geq 0$, because $c_1^2 \geq 0$, and
		\begin{align*}
		-4d_1b_1 & = -4 \left\lbrace \left( R_{max} - \tilde{R} \right) \sigma^2_Y \left( \mathring{\beta} - \tilde{\beta}\right) \sigma^2_X \right\rbrace \left\lbrace -\tau_X \left( \mathring{\beta} - \tilde{\beta}\right) \sigma^2_X \right\rbrace \\
		& = 4 \left( R_{max} - \tilde{R} \right) \sigma^4_X \sigma^2_Y \tau_X  \left( \mathring{\beta} - \tilde{\beta}\right)^2 \\
		& \geq 0
		\end{align*}
		where the last inequality follows because $R_{max} \geq \tilde{R}$.
	\end{proof}

	\subsubsection{Identified Sets are not Unique}\label{sec:oster-idset}
	The bounding sets for the `true' treatment effect, for instance reported in column~5 in Table~3 in \citet{oster-2019}are defined as $[ \tilde{\beta}, \beta^*(R_{max},\delta=1) ] $, where
	\[
	\beta^*(R_{max},\delta=1) = \tilde{\beta} - \text{root of the quadratic equation in (\ref{eq:quad})}.
	\]
	
	The implication of proposition~\ref{prop1} is that, in general, there will be two real roots of the quadratic equation in (\ref{eq:quad}). Hence, in general, there will be two values of the bias in the treatment effect, and hence two values of $\beta^*$. Without the extremely restrictive assumption that the discriminant of the quadratic equation in (\ref{eq:quad}) is identically equal to zero, it is not possible to arrive at a unique `identified' set when $\delta=1$. Since \citet[Section 4.1.1, Section 4.1.2, and Table 3]{oster-2019} uses the bias-adjusted treatment effect computed under the assumption of $\delta=1$ in constructing her `identified sets', it is not clear how one of these two sets are chosen.
	
	To be more specific, I follow \citet{oster-2019} and and construct identified sets by choosing $R_{max}=0.61$ for the regressions corresponding to the first three rows of Table~\ref{table:maternal-coeff} and $R_{max}=0.53$ for the regressions corresponding to the last two rows of Table~\ref{table:maternal-coeff}. I report the results from this exercise in Table~\ref{table:oster-idset} and follow the same row numbering as Table~3 in \citet{oster-2019}. Let us begin by noting, in column~1, the magnitude of the discriminant of the quadratic equation in (\ref{eq:quad}). We can see that the discriminant is always positive. Hence, in each case, there are two real roots. I use the first real root to define $\beta^*_1$ (column~2) and the second real root to define $\beta^*_2$ (column~3). The important conclusion to draw is that one \textit{cannot} get a uniquely identified set.\footnote{The root selection algorithm outlined in section~\ref{sec:cubic-algo} requires at least one set of coefficients to produce a unique real root. This will not work for the quadratic case because proposition~\ref{prop1} shows that we do not have a unique real root for \textit{any} set of coefficients.} 
	
	\begin{center}
		[Table~\ref{table:oster-idset} about here]
	\end{center}
	
	For instance, for the first row (where child IQ is the outcome and months of breastfeeding is the treatment), the first identified set is $[0.017, 0.375]$ and the second identified set is $[-0.034, 0.017]$. In row~1, Table~3, \citet{oster-2019} reports the second of these as \textit{the} identified set. But there is no reason offered for this choice. What is basis on which one can choose between the two different identified sets? The same problem affects all the five rows in Table~3, \citet{oster-2019}. No reason is given for choosing one over the other identified set. This is especially important because in \textit{three cases} out of five, the conclusion of the bounding analysis would change if one set was chosen rather than the other. For instance, in the case of the first row (where child IQ is the outcome and months of breastfeeding is the treatment), the first identified does not include zero; the second identified set does include zero. The same is true for the analysis represented by row 2 and row 5.
	
	If the quadratic in (\ref{eq:quad}) does not have a unique root for $\delta=1$ and $R_{max}=0.61$ (or $R_{max}=0.53$), then how can \citet[column 5, Table~3]{oster-2019} report one identified set? There seem to be two possibilities. First, it seems that she has taken recourse to Assumption~3 in her paper to generate a unique root of the quadratic equation. Assumption~3, in \citet{oster-2019}, states that the sign of the covariance between the treatment variable and the actual index of observables is the same as the sign of the covariance between the treatment variable and the predicted index of observables. The meaning and import of this assumption is explained thus.
	\begin{quote}
		Effectively, this assumes that the bias from the unobservables is not so large that it biases the \textit{direction} of the covariance between the observable index and the treatment. Under Assumption~3, if $\delta=1$, there is a unique solution \citep[pp.~194, emphasis in original]{oster-2019}.
	\end{quote}
	It is not clear how the sign restriction on the covariance between the treatment variable and the index of observables can generate a unique root of the quadratic equation in (\ref{eq:quad}). \citet{oster-2019} does not provide a proof of this important claim in the paper or in the appendix. 
	
	The second possibility is that among the two real roots of the quadratic equation, \citet{oster-2019} choose the one that implies a lower absolute magnitude of omitted variable bias. In each of the five cases reported in \citet[Table 3]{oster-2019}, one can see this by matching the identified sets that emerge from Table~\ref{table:oster-idset}. If this is the implicit reasoning behind the choice among the two real roots of the quadratic equation in (\ref{eq:quad}), then it needs to be justified on theoretical grounds. As it stands, it is unjustified and comes across as an \textit{ad hoc} assumption. Why should we assume that the omitted variables are such that they will produce the lower of the two possible magnitudes of bias?

	\subsection{Does $\delta^*$ Provide Useful Information?}\label{sec:deltastar}
	
	Let us consider the second strategy proposed in \citet{oster-2019} to deal with non-uniqueness, i.e. computing the value of $\delta$ (the relative selection on unobservables) that is consistent with a zero treatment effect, denoting this as $\delta^*$, and drawing conclusions about the problem of omitted variable bias by comparing $\delta^*$ with $1$. This strategy has at least three problems. First, it does not provide us with any identified \textit{set} of values of the true treatment effect, $\beta$. It only gives us one number, $\delta^*$. Second, in many cases, as I demonstrate below, $\delta^*$ can be extremely sensitive to the choice of $R_{max}$. Even a small error in choosing $R_{max}$ can greatly magnify $\delta^*$ and thereby lead to incorrect conclusions. Third, in some cases, the conclusions drawn from $\delta^*$ does not accord with the conclusions drawn on the basis of the bounding set that I have proposed above. When there is such a conflict, it seems better to use the bounding set than to rely on one number, $\delta^*$, because the former strategy is more robust. 
	
	\subsubsection{What is $\delta^*$?}
	Recall that $\delta^*$ is the degree of selection on unobservables that is consistent with zero treatment effect. If treatment effect is zero, i.e. $\beta=0$, then $\tilde{\beta}-\nu=0$. Hence, $\nu=\tilde{\beta}$. By plugging $\nu=\tilde{\beta}$ in (\ref{eq:cubic}), we get a relationship between $\delta$ and $R_{max}$ that can be expressed with the following function, 
	\begin{equation}\label{del-func}
	\delta = f(R_{max}) := \frac{C}{A\left( R_{max}-\tilde{R}\right) +B}, \quad \tilde{R} \leq R_{max} \leq 1,
	\end{equation}
	where
	\begin{align*}
	A = & \tilde{\beta}\sigma_Y^2(\sigma_X^2 - \tau_X) + \sigma_Y^2\sigma_X^2(\mathring{\beta} - \tilde{\beta}), \\
	B = & \tilde{\beta}^3(\sigma_X^2\tau_X - \tau_X^2) + \tilde{\beta}^2\sigma_X^2\tau_X(\mathring{\beta} - \tilde{\beta}),\\
	C = & \tilde{\beta}\sigma_Y^2\tau_X(\tilde{R} - \mathring{R}) + \tilde{\beta}\sigma_X^2\tau_X(\mathring{\beta} - \tilde{\beta})^2 \\
	 & \quad + \tilde{\beta}^3(\sigma_X^2 \tau_X - \tau_X^2) +  2\tilde{\beta}^2\sigma_X^2 \tau_X(\mathring{\beta} - \tilde{\beta}),
	\end{align*}
	are known constants that can be computed once we have estimated the short, intermediate and auxiliary regressions. When we plug in a value of $R_{max}$ in (\ref{del-func}), we get the corresponding $\delta^*$ by evaluating the function at that value of $R_{max}$.
	
	For the function in (\ref{del-func}) to be meaningful and useful, we need to impose some restrictions. First, if $\tilde{\beta}=0$, then $C=0$, and hence $\delta=0$ for all values of $R_{max}$. This is not interesting. So, we assume $\tilde{\beta} \neq 0$. Second, given that $\tilde{\beta} \neq 0$, if $A=0$, then $\delta$ is the constant function. It does not vary with $R_{max}$. Once again, this is not interesting for us. Hence, we impose the condition that $A \neq 0$. Third, the function is not defined at the point $R_{max}=R^*$, where $R^*=\tilde{R}-(B/A)$. Hence, we need to exclude this point from the domain of definition of the function. There are two cases to consider.
	
	\subsubsection{Case~1}
	In this case, $R^* \notin [ \tilde{R}, 1 ] $. Hence, the function $f$ is defined on every point in $[ \tilde{R}, 1 ]$. Moreover, it is differentiable on  $(\tilde{R}, 1)$ because it is a rational function and is defined on each point on this closed interval. The derivative of the function in (\ref{del-func}), on the open interval, is given by
	\begin{equation}\label{del-func-deriv}
	f'(R_{max}) = \frac{-AC}{ \left[  A\left( R_{max}-\tilde{R}\right) +B \right]^2}.
	\end{equation}
	\begin{proposition}
		The function in (\ref{del-func}) is monotone.
	\end{proposition}
	\begin{proof}
		If $C=0$, then $f'=0$ and $f$ is a constant function, i.e. both increasing and decreasing. If $C \neq 0$, we have either that $A$ and $C$ have the same sign or that they have opposite signs. If $A$ and $C$ are of the same sign, then using (\ref{del-func-deriv}), we see that $f'<0$. Hence, $f$ is strictly decreasing. If $A$ and $C$ are of opposite signs, then (\ref{del-func-deriv}) shows that $f'>0$ and $f$ is strictly increasing.  
	\end{proof}

\begin{figure}
	\centering
	\includegraphics[scale=0.65]{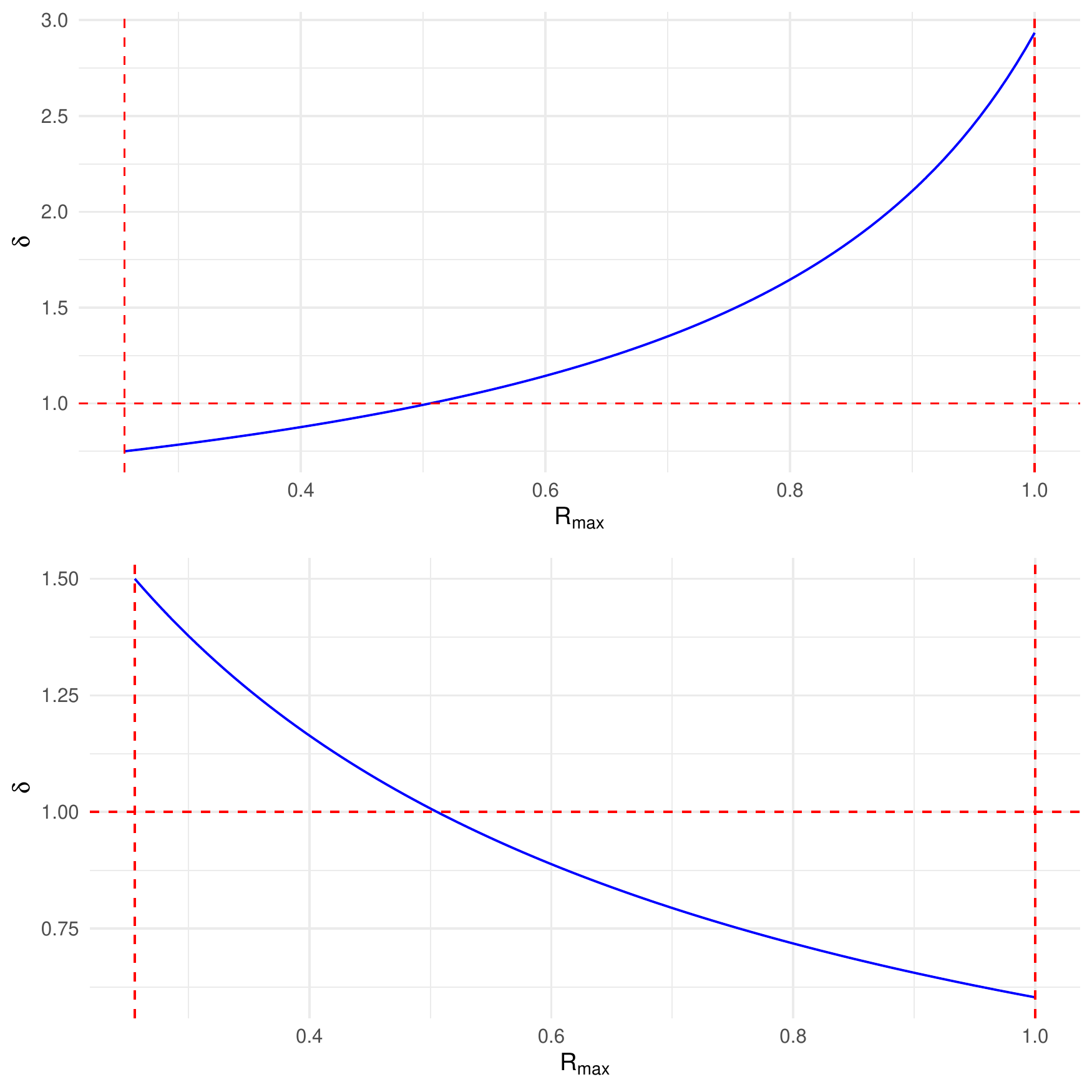}
	\caption{\textit{The figure plots two examples of the graph of the function, $\delta=f(R_{max}):=C/(A(R_{max}-\tilde{R})+B)$ over the domain, $\left[ \tilde{R}+0.01,1\right] $. For the top panel, $A=-2,B=2,C=1.5,\tilde{R}=0.256$; for the bottom panel, $A=4,B=2,C=3,\tilde{R}=0.256$. For a discussion, see section~\ref{sec:deltastar}.}}
	\label{fig:delplot1}
\end{figure}

	Two examples of the $f$ function are given in Figure~\ref{fig:delplot1}, one where its graph is upward sloping and another where its graph is downward sloping.
	
	\begin{center}
		[Figure~\ref{fig:delplot1} about here]
	\end{center}
	
	I give sufficient conditions for these two types of the $f$ function for the case when $R^* \notin [ \tilde{R}, 1 ] $ and provide some intuition for these conditions. 
	
	\begin{proposition}\label{neg-slope}
		If $0 < \tilde{\beta} <\mathring{\beta}$ or $ \mathring{\beta} < \tilde{\beta} < 0$, then the function in (\ref{del-func}) has a downward sloping graph.
	\end{proposition}
	\begin{proof}
		If $0 < \tilde{\beta} <\mathring{\beta}$ or $ \mathring{\beta} < \tilde{\beta} < 0$, then it can be easily verified that the parameters $A$ and $C$ are of the same sign. Using (\ref{del-func-deriv}), we get the result.
	\end{proof}
	Intuitively, what this sufficient condition says is this: if in moving from the short to the intermediate regression, the treatment effect moves towards zero without changing sign, then $\delta$ and $R_{max}$ have a negative relationship among themselves.
	
	\begin{proposition}\label{pos-slope}
		If either of the following two conditions are satisfied,
		\begin{enumerate}
			\item $\tilde{\beta} > (\sigma_X^2/\tau_X)\mathring{\beta}$, and 
			
			\begin{enumerate}
				\item $\tilde{\beta}^2 < (\sigma_X^2/\tau_X)\mathring{\beta}^2 + (\sigma_Y^2/\tau_X)(\tilde{R}-\mathring{R})$, if $\tilde{\beta}>0$, or
				
				\item $\tilde{\beta}^2 > (\sigma_X^2/\tau_X)\mathring{\beta}^2 + (\sigma_Y^2/\tau_X)(\tilde{R}-\mathring{R})$, if $\tilde{\beta}<0$,
			\end{enumerate}
			
			\item $\tilde{\beta} < (\sigma_X^2/\tau_X)\mathring{\beta}$, and 
			
			\begin{enumerate}
				\item $\tilde{\beta}^2 > (\sigma_X^2/\tau_X)\mathring{\beta}^2 + (\sigma_Y^2/\tau_X)(\tilde{R}-\mathring{R})$, if $\tilde{\beta}>0$, or
				
				\item $\tilde{\beta}^2 < (\sigma_X^2/\tau_X)\mathring{\beta}^2 + (\sigma_Y^2/\tau_X)(\tilde{R}-\mathring{R})$, if $\tilde{\beta}<0$,
			\end{enumerate}
			
		\end{enumerate}
		then the function in (\ref{del-func}) has a upward sloping graph.
	\end{proposition}
	\begin{proof}
		Note that $A<0$ iff $\tilde{\beta} > (\sigma_X^2/\tau_X)\mathring{\beta}$. Similarly, note that $C>0$ iff $ \tilde{\beta}\left[ \sigma_X^2 \tau_X\mathring{\beta}^2 + \sigma_Y^2 \tau_X(\tilde{R}-\mathring{R}) - \tau_X^2 \tilde{\beta}^2\right] <0 $. Now, using (\ref{del-func-deriv}), we get the result.
	\end{proof}
	One way in which this sufficient condition will be satisfied is this: if in moving from the short to the intermediate regression, the treatment effect changes sign and the difference of their absolute values is sufficiently large, then $\delta$ and $R_{max}$ has a positive relationship among themselves. As an example, consider $\tilde{\beta}=1$ and $\mathring{\beta}=-1.5$. Since $\sigma_X^2>\tau_X>0$, $\sigma_Y^2>0$ and $\tilde{R}>\mathring{R}$, this choice satisfies condition~1 (a) in proposition~\ref{pos-slope}. As another example, consider a scenario where $\tilde{\beta}=-1$ and $\mathring{\beta}=1.5$. Here, we can see that condition~2 (b) is satisfied.
	
	\textit{Interpretation of Slope.} When the graph of the function in (\ref{del-func}) is downward sloping, then this implies that for the treatment effect to be zero ($\beta=0$), a relatively high value of $R_{max}$ will be associated with a relatively low value of $\delta$. This can be interpreted in two different ways. On the one hand, this means that if the omitted variable is relatively more important in explaining the variation in the outcome variable than the included controls (high $R_{max}$), then only a small degree of selection on unobservables (low $\delta$) will ensure that the treatment effect is zero. If the degree of selection on unobservables is high, then the treatment effect is unlikely to be reduced to zero. On the other hand, it also means that if the degree of selection on unobservables is low, then only a relatively high importance of the omitted variable in explaining variation in the outcome variable compared to the included controls (high $R_{max}$) can ensure that the treatment effect is reduced to zero. If the omitted variable is relatively less important in explaining the variation in the outcome variable compared to the included controls, then the treatment effect cannot be washed out due to omitted variable bias. 
	
	On the other hand, when graph of the function in (\ref{del-func}) is upward sloping, exactly the opposite interpretation is valid. For the treatment effect to be zero ($ \beta=0 $), a relative high value of $R_{max}$ will be associated with a relatively high value of $\delta$. Once again, this can be interpreted in two different ways. On the one hand, this means that if the omitted variable is relatively more important in explaining the variation in the outcome variable (high $R_{max}$), then only a high degree of selection on unobservables than the included controls (high $\delta$) can ensure that the treatment effect is zero. A low degree of selection on unobservables will not reduce the treatment effect to zero. On the other, if the degree of selection on unobservable is low (low $\delta$) then only if the omitted variable is also relatively unimportant in explaining the variation of the outcome variable (low $R_{max}$), will the treatment effect be reduced to zero. If the omitted variable explains a relatively large part of the variation in the outcome variable, then the treatment effect cannot be reduced to zero due to omitted variable bias.
	
	Whatever interpretation we accord to $\delta^*$ in one case (negative slope) will have to be completely reversed in the other case (positive slope). Since we cannot \textit{a priori} rule out one or the other sign of the derivative of the function in (\ref{del-func}), if we use $\delta^*$ to draw conclusions about the severity or otherwise of the problem of omitted variable bias, our conclusion remains open to the need for a complete reversal of interpretation. 
	
	\subsubsection{Case~2}
	In this case, $R^* \in [ \tilde{R}, 1 ] $. Hence, the function is only defined on the intersection of two half-open intervals,
	\[
	\left\lbrace R_{max} | \tilde{R} \leq R_{max} < R^*\right\rbrace \cup  \left\lbrace  R_{max} | R^* < R_{max} \leq 1 \right\rbrace.
	\]
	The analysis of case~1 can now be applied to the two half-open intervals individually because the function is monotone on each of the half-open intervals. But there is an important implication of this case for using $\delta^*$ to draw conclusions about omitted variable bias. If a researcher computes the value of $\delta^*$ using (\ref{del-func}), compares it to $1$ and then draws conclusions about the problem of omitted variable bias, then, if this case holds, the researcher is likely to get very non-robust results. This is because the function is discontinuous at $R^*$. Hence, around $R^*$, the magnitude of $\delta^*$ is extremely sensitive to the choice of $R_{max}$. Even a slight error in choosing $R_{max}$ will greatly magnify the error in the magnitude of $\delta^*$.
	

	\subsubsection{$\delta^*$ Does not Match up with the Bounding Sets}
	In column~4 in Table~\ref{table:oster-idset}, I have reported the values of $\delta^*$ that was computed with (\ref{del-func}). In row~3, Table~\ref{table:oster-idset}, the value of $\delta^*$ is $1.36$. If we followed Oster's methodology, we would conclude that the reported estimate of the treatment effect is reliable, i.e. even after we take account of omitted variable bias, the true treatment effect is likely to remain different from zero. If we turn to the bounding sets reported in the third block of Table~\ref{table:bset-maternal}, we see that this conclusion is not wholly warranted. This is because, if $\delta>1$, the bounding set, computed according to my methodology, will include zero (row~12 in Table~\ref{table:bset-maternal}). For instance, if the omitted variables are relatively more important that the observed control variables in explaining the variation in the treatment variable (breastfeeding), then the relative degree of selection on unobservables would be larger than unity. In that case, the conclusion drawn on the basis of $\delta^*=1.36$, that omitted variable bias is not a problem, would be incorrect.
	
	If we look at row~4 in Table~\ref{table:oster-idset} and compare it with the penultimate block of results in Table~\ref{table:bset-maternal}, we will see the same problem. In row~4 in Table~\ref{table:oster-idset}, the value of $\delta^*$ is $1.08$. If we follow Oster's methodology, then we should conclude that omitted variable bias is not a serious problem. Now turn to row~16 in Table~\ref{table:bset-maternal}. Using the numbers in that row, we can see that, if $\delta>1$, the bounding set for the treatment effect is $[-2070, 118]$. Hence, the bounding set includes zero. Hence, if the omitted variables are relatively more important than the observed controls in explaining the variation in the treatment variable (drinking during pregnancy), then the relative degree of selection on unobservables might very well be larger than unity. In that case, the conclusion drawn on the basis of $\delta^*=1.08$, that omitted variable bias is a not problem, would be incorrect.   
	
	\subsubsection{Modified Procedure to Use $\delta^*$}
	The problem of discontinuity and non-correspondence with bounding sets suggests that the use of $\delta^*$ is fraught with problems. But, if a researcher must use it, then I would suggest a modified procedure. First, the researcher must ascertain whether $R^* \in [ \tilde{R}, 1 ] $, i.e. whether the point of discontinuity lies between $\tilde{R}$ and $1$. If the answer is yes, then the use of $\delta^*$ should be avoided. This is because the point of discontinuity lies in the interval of interest, $[ \tilde{R}, 1 ]$ and makes the analysis very unstable. Second, if $R^* \notin [ \tilde{R}, 1 ] $, then the researcher should ascertain the slope of the graph of the function in (\ref{del-func}). Since the interpretation is diametrically opposite depending on whether the sign is positive or negative, the researcher should note and report the sign of the derivative at any one point in the interval (monotonicity ensures that the sign of the derivative does not change). Third, the researcher can now report the value of $\delta^*$ and draw appropriate conclusion about omitted variable bias.
	
	In the last three columns of Table~\ref{table:oster-idset}, I have reported these three things for the five regression models I have studied in this paper. In each case, we can see that $R^* \notin [ \tilde{R}, 1 ] $. Hence, it is valid to carry out the $\delta^*$ analysis. We also see that in each case, the slope of the graph of the function in (\ref{del-func}) is negative. Hence, with all the caveats noted above, this perhaps allows us to interpret $\delta^*$ as done in \citet{oster-2019}. 
		
	\section{Concluding Comments}\label{sec:concl}
	Omitted variable bias is an ubiquitous problem in applied econometric work. Quantifying the magnitude of bias and computing bias-adjusted treatment effects is an important area of research. Building on earlier work by \citet{aet-2005}, in a recent contribution, \citet{oster-2019} has proposed a novel methodology to compute bias-adjusted treatment effect when there is proportional selection on observables and unobservables. In this paper, I have argued that while \citet{oster-2019} posed the problem correctly, her proposed solutions are problematic. I have instead proposed an alternative methodology to compute bounding sets for the true treatment effect. 
	
	My proposed methodology relies on two mathematical results. First, for a cubic equation, it is possible to use the discriminant to demarcate regions of the parameter space where a unique real root is guaranteed. Second, the roots of any polynomial are continuous functions of the coefficients. Using these two ideas, I have proposed an algorithm to compute real roots of the cubic equation and use them to construct an empirical distribution of the bias-adjusted treatment effect (BATE). Using this empirical distribution, one can construct approximate confidence intervals for the true treatment effect.
	
	To conclude the discussion, let me give a quick summary of the proposed methodology for the benefit of applied researchers.
	
		\begin{itemize}
			\item Estimate the short regression and store the coefficient on the treatment variable as $\mathring{\beta}$ and the R-squared as $\mathring{R}$.
			\item Estimate the intermediate regression and store the coefficient on the treatment variable as $\tilde{\beta}$ and the R-squared as $\tilde{R}$.
			\item Estimate an auxiliary regression by regressing the treatment variable on all the controls that were excluded from the short regression. Store the variance of the residual from this regression as $\tau_X$.
			\item Store the variance of the outcome variable as $\sigma^2_y$ and the variance of the control variable as $\sigma^2_X$.
			\item Form the cubic equation in (\ref{eq:cubic}). 
			\item Choose a bounded box in the ($\delta, R_{max}$) plane over which $\delta$ and $R_{max}$ can vary. Demarcate the URR and NURR regions in the box. 
			\item If the box is completely contained in the URR: Choose a $N \times N$ grid to cover the $URR$ area and solve the cubic at each point on the grid. Collect the $N^2 \times 1$ vector of real roots, $\nu$, of the cubic equation. This gives the empirical distribution of the treatment bias. Define $\beta^* = \tilde{\beta}-\nu$ and note that this is the bias-adjusted treatment effect. Use the empirical distribution of $\beta^*$ to define bounding sets for the `true' treatment effect.
			\item If the box is partly contained in the NURR: Compute roots on the URR area as outlined above. Using grid points in URR that reside on the boundary of URR and NURR, select the boundary points of the grid that lie in NURR, which are within a `small' distance of the former points. Compute the three real roots on the boundary points of the grid that lie in NURR and select the root that is closest in absolute value to the unique real root in the closest grid point in URR. Iterate this process using the previously selected roots until all grid points in NURR are exhausted. Define $\beta^* = \tilde{\beta}-\nu$ using all the selected real roots and note that this is the bias-adjusted treatment effect. Use the empirical distribution of $\beta^*$ to define bounding sets for the `true' treatment effect. 
			\item If the box is wholly contained in NURR: Extend the dimension of the box until there is non-empty intersection with URR, and then repeat the steps outlined in the previous case. 
		\end{itemize}
	
	The method outlined above will give bounding sets for the choice of the bounded box chosen by the researcher. It is important that a researcher draw on knowledge of the institutional details of the substantive issue under investigation in identifying the correct range for $\delta$ and $R_{max}$. For instance, if a particular research question has an omitted variable that is understood to be very important in explaining variation in the treatment variable, then it might be justified to use $\delta>1$. If, on the other hand, the researcher is sure that all important variables have been included in the model, and hence, that the omitted variable is relatively less important in explaining variation in the treatment variable, then a range of $\delta<1$ might be justified. Similar considerations should be used to infer a correct upper bound for $R_{max}$. The bounds generated for the BATE will only be as good as the choice of the bounded box chosen by the researcher.

	\bibliographystyle{apalike}
	\bibliography{biasrefs}
	

	\newpage
\begin{table}[hbt!]
	\begin{center}
		\begin{threeparttable}
			\caption{Parameter Estimates for the Analysis of Maternal Behavior on Child Outcomes}
			\label{table:maternal-coeff}
			\begin{tabular}{lcccccc}
				\toprule
				
				& \multicolumn{3}{c}{Short Regression} & \multicolumn{3}{c}{Intermediate Regression} \\
				\cline{2-7} \\[-1.8ex]
				 & (1) & (2) & (3) & (4) & (5) & (6) \\
				\cline{2-7} \\[-1.8ex]
				Outcome, Treatment & $\mathring{\beta}$ & Std Err & $\mathring{R}$ & $\tilde{\beta}$ & Std Err & $\tilde{R}$ \\ 
				\hline \\[-1.8ex] 
				
				IQ, Breastfeed & 0.044 & 0.003 & 0.045 & 0.017 & 0.002 & 0.256 \\ 
				IQ, Drink Preg & 0.176 & 0.026 & 0.008 & 0.050 & 0.023 & 0.249 \\ 
				IQ, LBW + Preterm & -0.188 & 0.057 & 0.004 & -0.125 & 0.050 & 0.251 \\ 
				BW, Smoke Preg & -183.115 & 12.933 & 0.319 & -172.51 & 13.285 & 0.352 \\ 
				BW, Drink Preg & -16.668 & 5.156 & 0.301 & -14.149 & 5.065 & 0.338 \\ 
				
				\bottomrule
			\end{tabular}
			\begin{tablenotes}
				\small
				\item \textit{Notes:} This table reports the parameter estimates from the short regression ($\mathring{\beta}$, Std Err, $\mathring{R}$) and the intermediate regressions ($\tilde{\beta}$, Std Err, $\tilde{R}$) and corresponds to \citet[Table~3]{oster-2019}. For some details of each of the models, see the first paragraph of section~\ref{sec:appl}.
			\end{tablenotes}
		\end{threeparttable}
	\end{center}
	
\end{table}

	\begin{sidewaystable}[hbt!]
		\begin{center}
			\begin{threeparttable}
				\caption{Empirical Distributions of Omitted Variable Bias and the Bias-Adjusted Treatment Effect (BATE) related to the Effect of Maternal Behavior on Child Outcomes}
				\label{table:bset-maternal}
				\begin{tabular}{lccccc}
					\toprule
					
					 & (1) & (2) & (3) & (4) & (5) \\
					\cline{2-6} \\[-1.8ex]
					& 2.5\% & 5\% & 50\% & 95\% & 97.5\% \\ 
					\hline \\[-1.8ex] 
					IQ, Breastfeed: Bias (Region 1) & $0$ & $0$ & $0.009$ & $0.034$ & $0.039$ \\ 
					IQ, Breastfeed: BATE (Region 1) & $$-$0.021$ & $$-$0.017$ & $0.009$ & $0.017$ & $0.017$ \\ 
					IQ, Breastfeed: Bias (Region 2) & $0$ & $0.004$ & $0.059$ & $0.119$ & $0.133$ \\ 
					IQ, Breastfeed: BATE (Region 2) & $$-$0.116$ & $$-$0.102$ & $$-$0.041$ & $0.013$ & $0.017$ \\  
					
					& & & & &\\
					IQ, Drink Preg: Bias (Region 1) & $0$ & $0.001$ & $0.036$ & $0.138$ & $0.155$ \\ 
					IQ, Drink Preg: BATE (Region 1) & $$-$0.104$ & $$-$0.087$ & $0.014$ & $0.049$ & $0.050$ \\ 
					IQ, Drink Preg: Bias (Region 2) & $0$ & $0.016$ & $0.221$ & $0.889$ & $1.158$ \\ 
					IQ, Drink Preg: BATE (Region 2) & $$-$1.107$ & $$-$0.839$ & $$-$0.171$ & $0.034$ & $0.050$ \\ 
					
					& & & & &\\
					IQ, LBW + Preterm: Bias (Region 1) & $$-$0.071$ & $$-$0.063$ & $$-$0.017$ & $$-$0.001$ & $0$ \\ 
					IQ, LBW + Preterm: BATE (Region 1) & $$-$0.125$ & $$-$0.124$ & $$-$0.108$ & $$-$0.062$ & $$-$0.054$ \\ 
					IQ, LBW + Preterm: Bias (Region 2) & $$-$0.300$ & $$-$0.272$ & $$-$0.097$ & $$-$0.007$ & $0$ \\ 
					IQ, LBW + Preterm: BATE (Region 2) & $$-$0.125$ & $$-$0.118$ & $$-$0.028$ & $0.147$ & $0.175$ \\

					& & & & &\\
					BW, Smoke Preg: Bias (Region 1) & $$-$151.390$ & $$-$120.190$ & $$-$18.521$ & $$-$0.421$ & $0$ \\ 
					BW, Smoke Preg: BATE (Region 1) & $$-$172.510$ & $$-$172.090$ & $$-$153.990$ & $$-$52.321$ & $$-$21.121$ \\ 
					BW, Smoke Preg: Bias (Region 2) & $$-$290.033$ & $$-$241.279$ & $217.838$ & $1,478.282$ & $1,897.682$ \\ 
					BW, Smoke Preg: BATE (Region 2) & $$-$2,070.193$ & $$-$1,650.792$ & $$-$390.348$ & $68.769$ & $117.522$ \\

					& & & & &\\
					BW, Drink Preg: Bias (Region 1) & $$-$17.044$ & $$-$15.014$ & $$-$3.587$ & $$-$0.098$ & $0$ \\ 
					BW, Drink Preg: BATE (Region 1) & $$-$14.149$ & $$-$14.051$ & $$-$10.562$ & $0.866$ & $2.895$ \\ 
					BW, Drink Preg: Bias (Region 2) & $$-$152.269$ & $$-$142.776$ & $$-$24.933$ & $$-$1.510$ & $0$ \\ 
					BW, Drink Preg: BATE (Region 2) & $$-$14.149$ & $$-$12.638$ & $10.784$ & $128.627$ & $138.120$ \\

					\bottomrule
				\end{tabular}
				\begin{tablenotes}
					\small
					\item \textit{Notes:} This table reports the empirical distributions of bias and the bias-adjusted treatment effect (BATE) for the five models resported in Table~\ref{table:bset-maternal}. Results about these models are discussed in \citet[Table 3]{oster-2019}. Quantiles have been computed using the algorithm discussed in section~\ref{sec:cubic-algo}.
				\end{tablenotes}
			\end{threeparttable}
		\end{center}
		
	\end{sidewaystable}

	\begin{table}[hbt!]
	\begin{center}
	\begin{threeparttable}
	\caption{Bounding Sets for the Treatment Effect for Different Step Sizes of Grid}
	\label{table:bdset-e}
	\begin{tabular}{lcccccc}
	\toprule
	
	 &  & \multicolumn{5}{c}{Quantiles of $\beta^*$} \\
	 \cline{3-7} \\[-1.8ex] 
	Step Size & Time (sec) & 2.5\% & 5\% & 50\% & 95\% & 97.5\%\\
	\hline \\[-1.8ex] 
	
	e=1/25 & $2.335$ & $$-$0.019$ & $$-$0.015$ & $0.010$ & $0.017$ & $0.017$ \\ 
	e=1/50 & $7.979$ & $$-$0.021$ & $$-$0.017$ & $0.009$ & $0.017$ & $0.017$ \\ 
	e=1/100 & $37.200$ & $$-$0.021$ & $$-$0.017$ & $0.009$ & $0.017$ & $0.017$ \\ 
	e=1/250 & $200.864$ & $$-$0.021$ & $$-$0.017$ & $0.009$ & $0.017$ & $0.017$ \\ 
	e=1/500 & $803.946$ & $$-$0.021$ & $$-$0.017$ & $0.008$ & $0.017$ & $0.017$ \\ 
	
	\bottomrule
\end{tabular}
\begin{tablenotes}
\small
\item \textit{Notes:} This table reports the bounding sets for the treatment effect in the regression reported in row~1, Table~\ref{table:maternal-coeff} as step size of the grid is reduced from $e=1/25$ to $e=1/500$. The estimation was carried out on a Lenovo Thinkpad 440s with a Intel Core i7 processor and 8GB of memory. 
\end{tablenotes}
\end{threeparttable}
\end{center}

\end{table}

	\newpage
	\begin{sidewaystable}[hbt!]
	\begin{center}
	\begin{threeparttable}
	\caption{Identified Set and $\delta^*$ Computed According to \cite{oster-2019}}
	\label{table:oster-idset}
	\begin{tabular}{lcccccc}
	\toprule
	
	 & (1) & (2) & (3) & (4) & (5) & (6) \\
	\cline{2-7} \\[-1.8ex]
	&  $D$  &  ID Set 1  & ID Set 2 & $\delta^*$ & Discont & Slope \\ 
	\hline \\[-1.8ex] 
	\underline{$R_{max}=0.61$} &  &  &  & & & \\ 
	
	IQ, Breastfeed & 20.32 & [0.017,0.375] & [-0.034,0.017] & 0.36 & FALSE & Negative \\ 
	IQ, Drink Preg & 0.002 & [0.050,8.410] & [-0.147,0.050] & 0.26 & FALSE & Negative \\ 
	IQ, LBW + Preterm & 0.0001 & [-79.874,-0.125] & [-0.125,-0.033] & 1.36 & FALSE & Negative \\ 
	
	&  &  &  & & & \\ 
	
	\underline{$R_{max}=0.53$} &  &  &  & & &  \\ 
	
	BW, Smoke Preg & 1905974 & [-3403.376,-172.511] & [-172.511,-49.713] & 1.16 & FALSE & Negative \\ 
	BW, Drink Preg & 267359453 & [-3615.821,-14.149] & [-14.149,0.944] & 0.94 & FALSE & Negative \\
	
	\bottomrule
\end{tabular}
\begin{tablenotes}
\small
\item \textit{Notes:} This table reports the identified set and $\delta^*$ computed according to the methodology in \citet[Table~3]{oster-2019}. $D$ stands for the discriminant of the quadratic equation in (\ref{eq:quad}). ID Set 1 is $[\tilde{\beta}, \tilde{\beta}-\nu_1]$ and ID Set 2 is $[\tilde{\beta}, \tilde{\beta}-\nu_2]$, where $\nu_1$ and $\nu_2$ are the two roots of (\ref{eq:quad}), respectively. Proposition~\ref{prop1} ensures that both these roots are real. $\delta^*$ is the value of $\delta$ that corresponds to $\beta=0$ and the relevant $R_{max}$ (which is $0.61$ for the first three rows and $0.53$ for the last two rows). $\delta^*$ has been computed with (\ref{del-func}). `Discont' is TRUE if $R^* \in [\tilde{R},1]$, and FALSE otherwise. `Slope' gives the slope the graph of the function $\delta=F(R_max)$ in (\ref{del-func}) on the domain $[\tilde{R},1]$; `Negative' denotes negative slope. For a discussion, see section~\ref{sec:oster-idset}.  
\end{tablenotes}
\end{threeparttable}
\end{center}

\end{sidewaystable}

	
\clearpage
\newpage

\clearpage
\newpage
\appendix
\counterwithin{figure}{section}

\section{Proof of Proposition~\ref{prop:3eqns}}\label{cubic-derive}
In this section of the appendix, I provide the details of the proof of Proposition~\ref{prop:3eqns} by deriving the equations in (\ref{eq:sol1}), (\ref{eq:sol2}) and (\ref{eq:sol3}), which can then be manipulated to derive the cubic equation in the omitted variable bias (\ref{eq:cubic}).

\subsection{Expression for Omitted Variable Bias}
\subsubsection{The Short Regression}
Consider the short regression,
\[
Y = \mathring{\alpha} + \mathring{\beta} X + \mathring{\varepsilon},
\]
the intermediate regression,
\[
Y = \tilde{\alpha} + \tilde{\beta} X + \sum_{j=1}^{J}\tilde{\psi}_j \omega^o_j + \tilde{\varepsilon},
\]
and the `true' model,
\[
Y = \alpha + \beta X + \sum_{j=1}^{J}\psi_j \omega^o_j + W_2 + \varepsilon.
\]
In the short regression, the $J \times 1$ vector of observable controls, $\omega^o$, and the scalar unobserved index, $W_2$, have been omitted. Hence, using the well-known formula for omitted variable bias (OVB) \citep{basu-2020}, we have 
\[
\mathring{\beta} \xrightarrow{p} \beta + \sum_{j=1}^{J} \psi_j \lambda_{\omega_j^o|X} + \lambda_{W_2|X},
\]
where $\lambda_{\omega_j^o|X}$ is the coefficient of $X$ in a regression of $\omega_j^o$ on $X$, and $\lambda_{W_2|X}$ is the coefficient of $X$ in a regression of $W_2$ on $X$, and $\xrightarrow{p}$ denotes convergence in probability as the sample size becomes large, i.e. $N \to \infty$. 

Note that,
\[
\lambda_{\omega_j^o|X} = \frac{\Cov(\omega_j^o,X)}{\Var(X)},
\]
so that
\[
\sum_{j=1}^{J} \psi_j \lambda_{\omega_j^o|X} = \sum_{j=1}^{J} \psi_j \frac{\Cov(\omega_j^o,X)}{\Var(X)} =   \frac{\Cov\left( \sum_{j=1}^{J}\psi_j\omega_j^o,X\right) }{\Var(X)} = \frac{\sigma_{1X}}{\sigma_X^2},
\]
because $W_1 = \sum_{j=1}^{J}\psi_j\omega_j^o$, and $\Cov(W_1,X)=\sigma_{1X}$. In a similar manner, we have
\[
\lambda_{W_2|X} =  \frac{\Cov(W_2,X)}{\Var(X)} = \frac{\sigma_{2X}}{\sigma_X^2}.
\]
Bringing this all together, we have
\begin{equation}\label{bs-1}
\mathring{\beta} \xrightarrow{p} \beta + \frac{\sigma_{1X}}{\sigma_X^2} + \frac{\sigma_{2X}}{\sigma_X^2}.
\end{equation}

\subsubsection{The Intermediate Regression}
Now consider the intermediate regression once again,
\begin{equation}\label{app-ireg1}
Y = \tilde{\alpha} + \tilde{\beta} X + \sum_{j=1}^{J}\tilde{\psi}_j \omega^o_j + \tilde{\varepsilon},
\end{equation}
and note that in this regression, the unobserved index, $W_2$, has been omitted. Let $\nu$ denote the bias in $\tilde{\beta}$ and let $\eta_j$ denote the bias in $\tilde{\psi}_j$. 

To find expressions for the OVB $\nu$ and $\eta_j$, define a linear projection of $W_2$ on $X$ and $\omega_1^o, \ldots, \omega_J^o$, as
\begin{equation}\label{app-lp1}
W_2 = \gamma_0 + \gamma_1 X  + \gamma_2 \omega_1^o + \cdots + \gamma_{J+1} \omega_J^o + r,
\end{equation}
where, by definition, the error term, $r$, is orthogonal to $X, \omega_1^o, \ldots, \omega_J^o$. Now, recall the hypothetical long regression, i.e. the true model,
\[
Y = \alpha + \beta X + \sum_{j=1}^{J}\psi_j \omega^o_j + W_2 + \varepsilon,
\]
and plug in the linear projection of $W_2$ in the true model to get
\begin{equation}\label{eq:int-lp}
Y = \left( \alpha + \gamma_0\right) + \left( \beta + \gamma_1 \right) X + \sum_{j=1}^{J}\left( \psi_j + \gamma_{j+1} \right) \omega^o_j + \left( r + \varepsilon\right).
\end{equation}
In this equation, the composite error term, $r+\varepsilon$, is orthogonal to all the regressors - because both $r$ and $\varepsilon$ are orthogonal. Hence, the probability limits of the parameters in the intermediate regression (\ref{app-ireg1}) can be read off from the corresponding coefficients in the above equation \citep[pp.~61-62]{wooldridge}. Comparing coefficients in (\ref{app-ireg1}) and (\ref{eq:int-lp}), we see that
\[
\tilde{\beta} \xrightarrow{p} \beta + \gamma_1,
\]
and
\[
\tilde{\psi}_j \xrightarrow{p} \psi + \gamma_{j+1}.
\]
Thus, once we find the coefficients in the linear projection (\ref{app-lp1}), we will be able to derive the expressions for the OVB of the parameters in the intermediate regression.

Consider the linear projection (\ref{app-lp1}) once again and let $\gamma$ be the $(J+1) \times 1$ vector of coefficients on the independent variables (excluding the constant), 
\[
\gamma = \begin{bmatrix}
\gamma_1 \\
\gamma_2 \\
\vdots \\
\gamma_{J+1}
\end{bmatrix}, 
\]
and let $W$ denote the $N \times (J+1)$ matrix of regressors (excluding the constant), 
\[
W = \begin{bmatrix}
X & \omega_{1}^o & \omega_{2}^o & \ldots & \omega_{J}^o  
\end{bmatrix},
\]
where $X$ is the $N \times 1$ vector of the treatment variable and, for $j=1,2, \ldots, J$, $\omega_{j}^o$ is the $N \times 1$ vector of the $j$-th observed control variable. From \citet[pp.~25]{wooldridge}, we have,
\begin{equation}\label{app-lp2}
\gamma = \left[ \Var (W) \right]^{-1} \Cov(W,W_2). 
\end{equation} 
Let $\omega_{jj}^o$ denote the variance of $\omega_j^o$, let $\omega_{jk}^o$ denote the covariance between $\omega_j^o$ and $\omega_k^o$, and let $\omega_{jx}^o$ denote the covariance between $\omega_j^o$ and $X$. Then, the $(J+1) \times (J+1)$ variance matrix is given by, 
\[
\Var (W) = \begin{bmatrix}
\sigma_X^2 & \omega_{1x}^o & \omega_{2x}^o & \ldots & \omega_{Jx}^o \\
\omega_{1x}^o & \omega_{11}^o & \omega_{12}^o & \ldots & \omega_{1 J}^o \\
\omega_{2x}^o & \omega_{21}^o & \omega_{22}^o & \ldots & \omega_{2 J}^o \\
\vdots & \vdots & & \ddots \\
\omega_{Jx}^o & \omega_{J1}^o & \omega_{J2}^o & \ldots & \omega_{JJ}^o 
\end{bmatrix}.
\]
A maintained assumption in this analysis is that \textit{the elements of the observed vector of controls is orthogonal to each other}.\footnote{See Proposition~\ref{prop:3eqns} in this paper and \citet[pp. 192]{oster-2019}.} This implies that, for $j,k=1, \ldots, J$, and $j \neq k$, $\omega_{jk}^o=0$. Thus, the variance matrix of $W$ simplifies to what is known as an arrowhead matrix (with nonzero elements on the first row, the first column and the principal diagonal),
\[
\Var (W) = \begin{bmatrix}
\sigma_X^2 & \omega_{1x}^o & \omega_{2x}^o & \ldots & \omega_{Jx}^o \\
\omega_{1x}^o & \omega_{11}^o & 0 & \ldots & 0 \\
\omega_{2x}^o & 0 & \omega_{22}^o & \ldots & 0 \\
\vdots & \vdots & & \ddots \\
\omega_{Jx}^o & 0 & 0 & \ldots & \omega_{JJ}^o 
\end{bmatrix}.
\]
This matrix structure is convenient because we will be able to easily invert it using existing results in applied linear algebra \citep{clarke_2019}. This is precisely where the first assumption stated in Proposition~\ref{prop:3eqns} is used.

Let us also note the second assumption of the analysis: $W_2$ \textit{is orthogonal to all the observed controls}.\footnote{See Proposition~\ref{prop:3eqns} in this paper and \citet[pp. 192]{oster-2019}.} This implies that the $(J+1) \times 1$ covariance vector is given by,
\[
\Cov (W_2,W) = \begin{bmatrix}
\sigma_{2X} \\
0 \\
0 \\
\vdots \\
0 
\end{bmatrix}.
\]
This is again a convenient vector, and will simplify the algebra considerably, because all elements other than the first one is zero. It is precisely to ensure that $\Cov (W_2,W)$ has this precise structure that the second assumption stated in Proposition~\ref{prop:3eqns} is used.

To compute the inverse of $\Var(W)$, we will draw on a result from \citet[appendix, pp.~2]{clarke_2019} regarding arrowhead matrices.\footnote{\citet{clarke_2019} borrows the result from \citet{najafi_etal_2014}.} Let us write $\Var(W)$ as
\[
\Var(W) = \begin{bmatrix}
a & z' \\
z & D
\end{bmatrix}
\]
where $a=\sigma_X^2$,
\[
z' = \begin{bmatrix}
\omega_{1x}^o & \omega_{2x}^o & \ldots & \omega_{Jx}^o
\end{bmatrix},
\]
and
\[
D = \begin{bmatrix}
\omega_{11}^o & 0 & \ldots & 0 \\
0 & \omega_{22}^o & \ldots & 0 \\
& & \ddots & \\
0 & 0 & \ldots & \omega_{JJ}^o
\end{bmatrix}.
\]
Then, the inverse of the variance matrix is given by
\begin{equation}\label{winv}
[\Var(W)]^{-1} = \begin{bmatrix}
0 & 0' \\
0 & D^{-1}
\end{bmatrix} + \rho v v',
\end{equation}
where
\[
v = \begin{bmatrix}
-1 & D^{-1} z
\end{bmatrix}',
\]
and
\[
\rho = \frac{1}{a - z' D^{-1} z}.
\]
Note that
\[
a - z' D^{-1} z = \sigma_X^2 - \sum_{j=1}^{J} \frac{\left( \omega_{jx}^o\right)^2 }{\omega_{jj}^o} = \tau_X,
\]
where the last equality comes from the fact that $\tau_X$ is the variance of the residual from a regression of $X$ on $\omega_1^o, \ldots, \omega_J^o$, and for all $j, k$, we have the maintained assumption that $\Cov(\omega_j^o, \omega_k^o)=0$. 

To see this, consider a regression of $X$ on $\omega_1^o, \ldots, \omega_J^o$ and write the predicted value from the regression as,
\[
\widehat{X} = \mu_1 \omega_1^o + \cdots + \mu_J \omega_J^o,
\]
and note that the variance of the residual is given by, 
\[
\tau_X = \Var(X-\widehat{X})=\sigma_X^2 + \Var(\widehat{X})-2\Cov(X,\widehat{X}).
\]
Since, for all $j, k = 1, 2, \ldots, J$, $j \neq k$, $\Cov(\omega_j^o, \omega_k^o)=0$, we have,
\[
\Var(\widehat{X}) = \sum_{j=1}^{J} \mu^2_j \omega^o_{jj}. 
\]
Moreover,
\[
\Cov(X,\widehat{X}) = \Cov(X, \sum_{j=1}^{J}\mu_j \omega_j^o) = \sum_{j=1}^{J} \mu_j \omega_{jx}^o.
\]
Since, for all $j, k = 1, 2, \ldots, J$, $j \neq k$, $\Cov(\omega_j^o, \omega_k^o)=0$, the coefficient on $\omega_{j}^o$, $\mu_j$, in the regression of $X$ on $\omega_1^o, \ldots, \omega_J^o$, is the same as would arise from a bivariate regression of $X$ on $\omega_{j}^o$. Hence, for $j=1, 2, \ldots, J$,
\[
\mu_j = \frac{\Cov(X,\omega_{j}^o)}{\Var(\omega_{j}^o)} = \frac{\omega_{jx}^o}{\omega_{jj}}.
\]
Hence,
\[
\tau_X = \sigma_X^2 + \sum_{j=1}^{J} \left( \frac{\omega_{jx}^o}{\omega_{jj}}\right)^2  \omega^o_{jj} - 2 \sum_{j=1}^{J} \frac{\omega_{jx}^o}{\omega_{jj}} \omega_{jx}^o =  \sigma_X^2 - \sum_{j=1}^{J} \frac{\left( \omega_{jx}^o\right)^2 }{\omega_{jj}^o}.
\]

Returning to (\ref{winv}), we see that
\begin{align*}
[\Var(W)]^{-1} & = \begin{bmatrix}
0 & 0 & 0 & \ldots & 0 \\
0 & \frac{1}{\omega_{11}^o}  & 0 & \ldots & 0 \\
\vdots & \vdots & & \ddots & 0 \\
0 & 0 &  & \ldots & \frac{1}{\omega_{JJ}^o}  
\end{bmatrix} + \\
& \frac{1}{\tau_X} \begin{bmatrix}
1 & -\frac{\omega_{1x}^o}{ \omega_{11}^o} & -\frac{\omega_{2x}^o}{\omega_{22}^o} & \ldots & -\frac{\omega_{Jx}^o}{ \omega_{JJ}^o} \\
-\frac{\omega_{1x}^o}{\omega_{11}^o} & \left( \frac{\omega_{1x}^o}{\omega_{11}^o} \right)^2  & \frac{\omega_{1x}^o \omega_{2x}^o}{\omega_{11}^o \omega_{22}^o} & \ldots &  \frac{\omega_{1x}^o \omega_{jx}^o}{\omega_{11}^o \omega_{JJ}^o} \\
\vdots & \vdots & \vdots & \ddots & \vdots \\
-\frac{\omega_{Jx}^o}{\omega_{JJ}^o} & \frac{\omega_{Jx}^o \omega_{1x}^o}{\omega_{JJ}^o \omega_{11}^o} & \frac{\omega_{Jx}^o \omega_{2x}^o}{\omega_{JJ}^o \omega_{22}^o} & \ldots &  \left( \frac{\omega_{Jx}^o}{\omega_{JJ}^o} \right)^2
\end{bmatrix},
\end{align*}
Using the expressions for $[\Var(W)]^{-1}$ and $\Cov(W,W_2)$, we get
\[
\begin{bmatrix}
\gamma_1 \\
\gamma_2 \\
\vdots \\
\gamma_{J+1}
\end{bmatrix} = \begin{bmatrix}
\frac{1}{\tau_X} & -\frac{\omega_{1x}^o}{\tau_X \omega_{11}^o} & -\frac{\omega_{2x}^o}{\tau_X \omega_{22}^o} & \ldots & -\frac{\omega_{Jx}^o}{\tau_X \omega_{JJ}^o} \\
-\frac{\omega_{1x}^o}{\tau_X \omega_{11}^o} &  &  & \ldots &  \\
\vdots & \vdots & & \ddots \\
-\frac{\omega_{Jx}^o}{\tau_X \omega_{JJ}^o} &  &  & \ldots &  
\end{bmatrix} \begin{bmatrix}
\sigma_{2X} \\
0 \\
0 \\
\vdots \\
0 
\end{bmatrix},
\]
where I have explicitly written out only the elements in the first row and column of $\left[ \Var(W)\right]^{-1} $ because only those elements will be necessary for our calculation. 

Multiplying out the matrices on the right, we see that the omitted variable bias in $\tilde{\beta}$ in the intermediate regression is given by
\begin{equation}
\gamma_1 = \frac{\sigma_{2X}}{\tau_X},
\end{equation}  
and the omitted variable bias in the coefficient on the $j$-th observed control, for $j=1, 2, \ldots, J$, is given by
\begin{equation}\label{bint-1}
\gamma_{j+1} = -\frac{\omega_{jx}^o}{\tau_X \omega_{jj}}.
\end{equation}  
It is important to note that if we did not rely on the two orthogonality assumptions stated in Proposition~\ref{prop:3eqns}, then we would not have arrived at these expressions. Without the pairwise orthogonality between the observed controls, we would not have been able to explicitly write out the elements of $[\Var(W)]^{-1}$, and without the orthogonality between the unobserved confounder and each observed control, we would have to deal with potentially many more nonzero elements in $\Cov(W,W_2)$.

\subsubsection{Bringing the Two Together}
Since the omitted variable bias in $\tilde{\beta}$ in the intermediate regression is denoted as $\nu$, we have, 
\begin{equation}\label{bint-2}
\nu = \frac{\sigma_{2X}}{\tau_X},
\end{equation}  
so that
\begin{equation}\label{bint-3}
\tilde{\beta} \xrightarrow{p} \beta + \nu. 
\end{equation}
Furthermore, using (\ref{bint-2}) in (\ref{bs-1}), we also get
\begin{equation}\label{bs-2}
\mathring{\beta} \xrightarrow{p} \beta + \frac{\sigma_{1X}}{\sigma_X^2} + \frac{\nu \tau_X}{\sigma_X^2}.
\end{equation}
Taking the difference between (\ref{bs-2}) and (\ref{bint-3}) gives us
\begin{equation}\label{eq:sol1-app}
\left( \mathring{\beta} - \tilde{\beta}\right) = \frac{\sigma_{1X}}{\sigma_X^2} - \nu \left( \frac{\sigma_X^2 - \tau_X}{\sigma_X^2}\right), 
\end{equation}
which is the equation in (\ref{eq:sol1}).

\subsection{Expressions for R-Squared}
\subsubsection{Short Regression}
Consider the short regression once again,
\[
Y_i = \mathring{\alpha} + \mathring{\beta} X_i + \mathring{\varepsilon}_i,
\]
and re-write it in deviation-from-mean form,
\[
\widetilde{Y}_i = \mathring{\beta} \widetilde{X}_i + \tilde{\mathring{\varepsilon}}_i,
\]
where $\widetilde{Y}_i=Y_i-\bar{Y}$ and $\widetilde{X}_i=X_i-\bar{X}$.  If we denote by $\mathring{R}$, the R-squared from this regression, then we have
\[
\mathring{R} = \frac{\sum_{i=1}^{N} \mathring{\beta}^2 \widetilde{X}_i^2}{ \sum_{i=1}^{N} \widetilde{Y}_i^2}.
\]
On dividing the numerator and denominator of the right-hand side by $N$ and then taking probability limits, we get, on rearranging,
\begin{equation}\label{r2-short}
\mathring{R} \sigma_Y^2 = \left( \beta + \frac{\sigma_{1X}}{\sigma_X^2} + \frac{\nu \tau_X}{\sigma_X^2}\right)^2 \sigma_X^2, 
\end{equation}
where we have used the fact that 
\[
\mathring{\beta} = \beta + \frac{\sigma_{1X}}{\sigma_X^2} + \frac{\nu \tau_X}{\sigma_X^2}.
\]

\subsubsection{Intermediate Regression}
Consider the intermediate regression once again,
\[
Y = \tilde{\alpha} + \tilde{\beta} X + \sum_{j=1}^{J}\tilde{\psi}_j \omega^o_j + \tilde{\varepsilon},
\]
and re-write it in deviation-from-mean form,
\[
\widetilde{Y} = \tilde{\beta} \widetilde{X} + \sum_{j=1}^{J}\tilde{\psi}_j \widetilde{\omega}^o_j + \tilde{\varepsilon}.
\]
If we denote by $\tilde{R}$, the R-squared from the intermediate regression, then we have, 
\[
\tilde{R} = \frac{\sum_{i=1}^{N} \tilde{\beta}^2 \widetilde{X}_i^2 + \sum_{j=1}^{J} \sum_{i=1}^{N} \tilde{\psi}^2_j \left( \widetilde{\omega}^o_{ji}\right)^2 + 2  \sum_{j=1}^{J} \sum_{i=1}^{N} \tilde{\beta} \widetilde{X}_i \tilde{\psi}_j \widetilde{\omega}^o_{ji} }{ \sum_{i=1}^{N} \widetilde{Y}_i^2}.
\]
On dividing the numerator and denominator of the right-hand side by $N$ and then taking probability limits, we get, on rearranging,
\begin{align*}
\tilde{R} \sigma_Y^2 = \tilde{\beta}^2 \sigma_{X}^2 + \sum_{j=1}^{J} \tilde{\psi}^2_j \omega_{jj}^o + 2 \tilde{\beta} \sum_{j=1}^{J} \tilde{\psi}_j \omega_{jx}^o.
\end{align*}
Since $\tilde{\beta} = \beta + \nu$, and 
\[
\tilde{\psi}_j = \psi_j - \frac{\omega_{jx}^o}{\tau_X \omega_{jj}},
\]
we get
\begin{align*}
\tilde{R} \sigma_Y^2 & = \left( \beta + \nu\right)^2 \sigma_{X}^2 + \sum_{j=1}^{J} \left( \psi_j - \frac{\omega_{jx}^o}{\tau_X \omega_{jj}}\right)^2 \omega_{jj}^o + \\
& \qquad 2 \left( \beta + \nu\right) \sum_{j=1}^{J} \left( \psi_j - \frac{\omega_{jx}^o}{\tau_X \omega_{jj}}\right) \omega_{jx}^o.
\end{align*}
Note that
\[
\sigma_1^2 = \Var(W_1) = \sum_{j=1}^{J} \psi_j^2 \omega_{jj}^o
\]
and 
\[
\sigma_{1X} = \Cov(W_1,X) = \sum_{j=1}^{J} \psi_j \omega_{jx}^o
\]
and, as we have seen above,
\[
\tau_X =  \sigma_X^2 - \sum_{j=1}^{J} \frac{\left( \omega_{jx}^o\right)^2 }{\omega_{jj}^o}.
\]
Using these relationships and simplifying, we get
\begin{equation}\label{r2-int}
\tilde{R} \sigma_Y^2 = \beta^2 \sigma_{X}^2 + \sigma_1^2 + \nu^2 \tau_X + 2 \beta \nu \tau_X + 2 \beta \sigma_{1X}.
\end{equation}

\subsubsection{Hypothetical Long Regression}
Finally, consider the hypothetical long regression,
\[
Y = \alpha + \beta X + \sum_{j=1}^{J}\psi_j \omega_j^o + W_2 + \varepsilon,
\]
and re-write it in deviation-from-mean form,
\[
\widetilde{Y} = \beta \widetilde{X} + \sum_{j=1}^{J} \psi_j \widetilde{\omega}_j^o + \widetilde{W}_2 + \widetilde{\varepsilon}.
\]
Denoting by $R_{max}$, the R-squared from this regression and using the same arguments as above, we get
\begin{equation}\label{r2-long}
R_{max} \sigma_Y^2 = \beta^2 \sigma_{X}^2 + \sigma_1^2 + \sigma_2^2 + 2 \beta \nu \tau_X + 2 \beta \sigma_{1X}.
\end{equation}

\subsubsection{Bringing it All Together}
Taking the difference between (\ref{r2-int}) and (\ref{r2-short}) gives us equation (\ref{eq:sol2}) in the main text, 
\[
\left( \tilde{R} - \mathring{R}\right) \sigma_y^2 = \sigma_1^2 + \tau_X \nu^2 - \frac{1}{\sigma_X^2}\left( \sigma_{1X} + \nu \tau_X\right)^2,
\]
and the difference between (\ref{r2-long}) and (\ref{r2-int}) gives us equation (\ref{eq:sol3}) in the main text,
\[
\left(R_{max} - \tilde{R}\right) \sigma_y^2 = \nu \left( \frac{\sigma_1^2 \tau_X}{\delta \sigma_{1X}} - \nu \tau_X\right).
\]

\section{Unique Real Root of a Cubic Equation}\label{cubic-solve}
Solving cubic equations is common in the engineering literature and for this presentation I draw partly on \citet[Appendix~1]{hellesland-etal-2013}. Consider the cubic equation in $t$,
\begin{equation}\label{cubic-sol}
a t^3 + b t^2 + c t + d = 0,
\end{equation}
where $a\neq 0$. Divide through by $a$ to get
\begin{equation}
t^3 + \frac{b}{a} t^2 + \frac{c}{a} t + \frac{d}{a}=0.
\end{equation}
A change of variable,
\[
x = t + \frac{b}{3a},
\]
can convert this into a `depressed' cubic,
\begin{equation}\label{depcub}
x^3 + px + q = 0,
\end{equation}
where, 
\begin{align*}
p & = \frac{3ac - b^2}{3 a^2} \\
q & = \frac{27 a^2 d + 2 b^3 - 9 abc}{27 a^3}.
\end{align*}
To solve (\ref{depcub}), we will express $x$ as the difference of two numbers, i.e. $x = a-b$. Since,
\begin{equation}\label{a3b3}
\left( a-b\right)^3 + 3 ab \left( a-b\right)- \left( a^3 - b^3\right) =0, 
\end{equation} 
we will get back (\ref{depcub})	from (\ref{a3b3}), where $x=a-b$, if the following two conditions are satisfied: 
\begin{equation}\label{ab}
ab=\frac{p}{3} 
\end{equation}
and
\begin{equation}\label{ab3}
a^3-b^3=-q.
\end{equation}
Thus, if we are able to solve for $a$ and $b$ in terms of $p$ and $q$, we will be able to get $x=a-b$, and from that we will be able to finally get the value of $t=x-(b/3a)$.

Note that the above two conditions, (\ref{ab}) and (\ref{ab3}), show that the sum and product of $a^3$ and $(-b)^3$ are $-q$ and $-p^3/27$, respectively. But this means that $a^3$ and $(-b)^3$ are the roots of the following quadratic equation in $y$,

\begin{equation}
y^2 + q y -  \frac{p^3}{27} = 0. 
\end{equation}
Denoting one of the roots of the quadratic as $a^3$, we have,
\[
a^3 = -\frac{q}{2} + \sqrt{\frac{q^2}{4}+\frac{p^3}{27}} = U_1
\]
and denoting the other root as $-b^3$, we get 
\[
-b^3 = -\frac{q}{2} - \sqrt{\frac{q^2}{4}+\frac{p^3}{27}},
\]
so that
\[
b^3 = \frac{q}{2} + \sqrt{\frac{q^2}{4}+\frac{p^3}{27}} = U_2.
\]
The solutions of the cubic equation (\ref{cubic-sol}) will depend on the sign of the discriminant
\begin{equation}\label{discrim}
D_3 = \frac{q^2}{4}+\frac{p^3}{27} = \frac{27q^2+4p^3}{108}.
\end{equation}

\begin{proposition}\label{prop:cubic-r1}
	If $D_3>0$, then the cubic equation has one real root and two complex roots. The unique real root is given by
	\[
	t_1 = \sqrt[3]{-\frac{q}{2} + \sqrt{\frac{q^2}{4}+\frac{p^3}{27}}} - \sqrt[3]{\frac{q}{2} + \sqrt{\frac{q^2}{4}+\frac{p^3}{27}}} - \frac{b}{3a},
	\]
	and the complex roots are given by
	\[
	t_2 = \omega \sqrt[3]{-\frac{q}{2} + \sqrt{\frac{q^2}{4}+\frac{p^3}{27}}} - \omega^2 \sqrt[3]{\frac{q}{2} + \sqrt{\frac{q^2}{4}+\frac{p^3}{27}}} - \frac{b}{3a},
	\]
	and
	\[
	t_3 = \omega^2 \sqrt[3]{-\frac{q}{2} + \sqrt{\frac{q^2}{4}+\frac{p^3}{27}}} - \omega \sqrt[3]{\frac{q}{2} + \sqrt{\frac{q^2}{4}+\frac{p^3}{27}}} - \frac{b}{3a},
	\]
	where $\omega$ is the cube root of unity given by
	\[
	\omega =e^{i \frac{2\pi}{3}}= -\frac{1}{2} + \frac{\sqrt{2}}{3} i
	\]
	and $i = \sqrt{-1}$.
\end{proposition}
\begin{proof}
	To see this, note that the possible pairs of $\left( a,b\right) $ that will satisfy (\ref{ab}) and (\ref{ab3}) are
	\[
	\left(\sqrt[3]{U_1},\sqrt[3]{U_2}\right), \left( \omega\sqrt[3]{U_1}, \omega^2\sqrt[3]{U_2}\right), \left(\omega^2\sqrt[3]{U_1}, \omega\sqrt[3]{U_2}\right). 
	\]
	Since $D_3>0$, $U_1$ and $U_2$ are real numbers. Hence, the unique real value of $x$ is given by
	\[
	x = a-b = \sqrt[3]{U_1} - \sqrt[3]{U_2} = \sqrt[3]{-\frac{q}{2} + \sqrt{\frac{q^2}{4}+\frac{p^3}{27}}} - \sqrt[3]{\frac{q}{2} + \sqrt{\frac{q^2}{4}+\frac{p^3}{27}}}
	\]
	and the corresponding unique real root of the original cubic equation (\ref{cubic-sol}) is given by
	\[
	t = \sqrt[3]{-\frac{q}{2} + \sqrt{\frac{q^2}{4}+\frac{p^3}{27}}} - \sqrt[3]{\frac{q}{2} + \sqrt{\frac{q^2}{4}+\frac{p^3}{27}}} - \frac{b}{3a}.
	\]
	The other two roots will be complex conjugate numbers because they involve $\omega$.
\end{proof} 

An immediate corollary follows. Cubic equations with real coefficients can have either one or three real roots. Complex roots occur in conjugate pairs. Thus, when the cubic has only real roots, i.e. three real roots, it will be the case that $D_3 \leq 0$.

\section{Figures}
In this section of the appendix, I present region plots and contour plots of the bias for each of the five models discussed in the main text of the paper. Parameter estimates and bounding sets of the treatment effect of these models are reported in Table~\ref{table:maternal-coeff} and Table~\ref{table:bset-maternal}, respectively.

\begin{figure}
	\centering
	\includegraphics[scale=0.65]{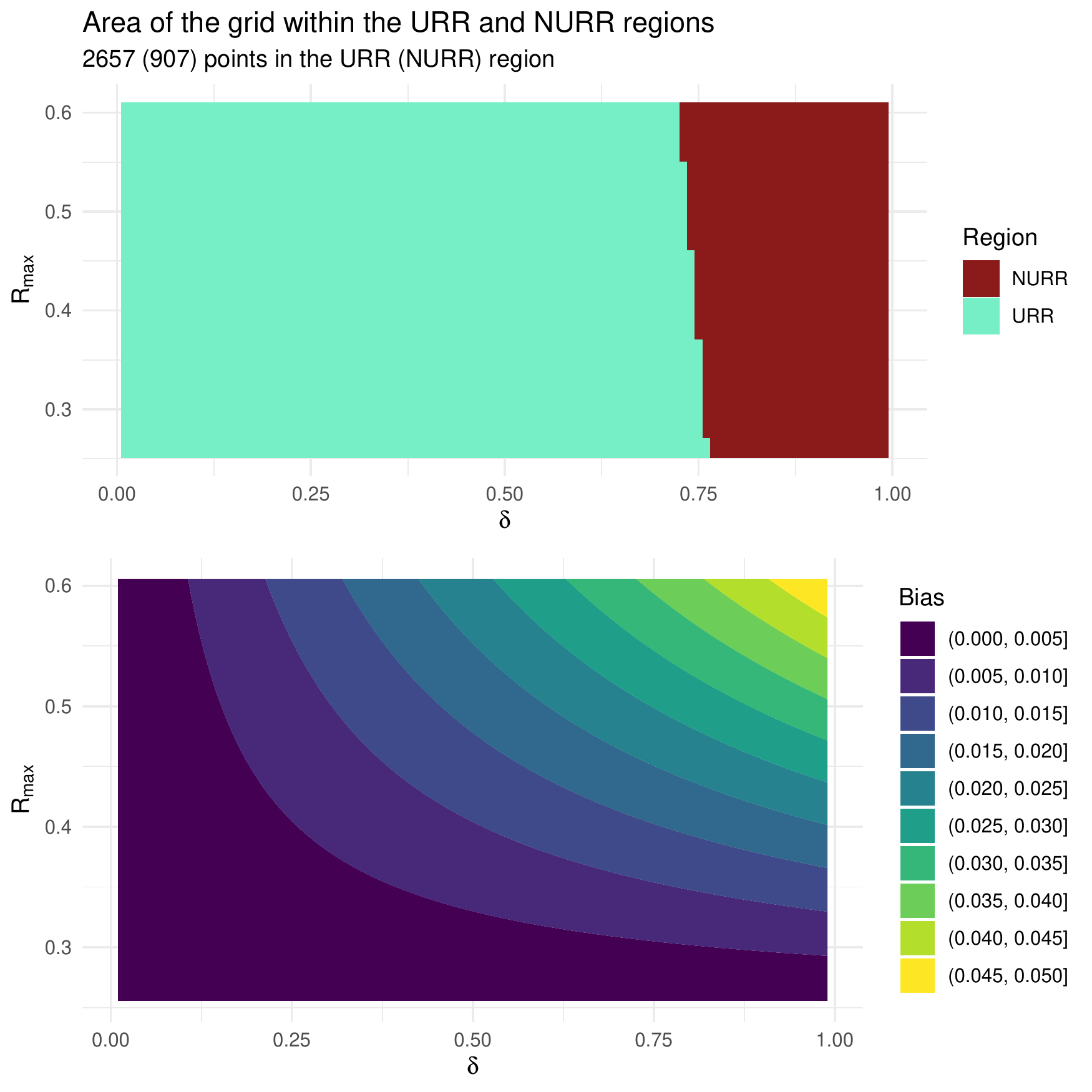}
	\caption{\textit{The figure identifies the region of unique real root (URR, top panel) and plots the contour of omitted variable bias (bottom panel) for the regression reported in row~1, Table~\ref{table:bset-maternal}. The bounded box is defined by $\delta \in [0.01,0.99]$, $R_{max} \in [\tilde{R},0.61]$, where $\tilde{R}=0.256$. Step size of grid = 0.01 in both the horizontal and vertical directions.}}
	\label{fig:mat11}
\end{figure}

\begin{figure}
	\centering
	\includegraphics[scale=0.65]{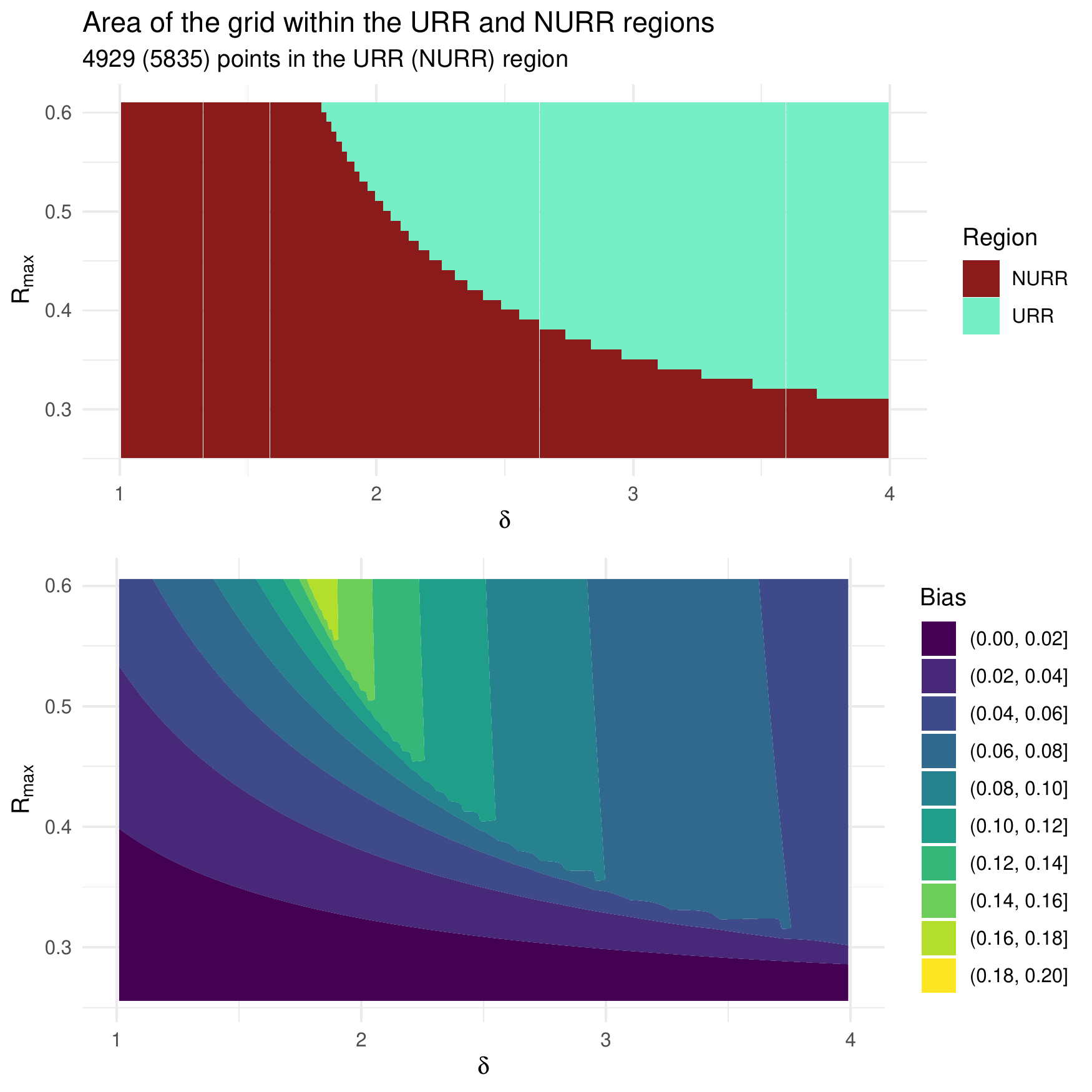}
	\caption{\textit{The figure identifies the region of unique real root (URR, top panel) and plots the contour of omitted variable bias (bottom panel) for the regression reported in row~1, Table~\ref{table:bset-maternal}.The bounded box is defined by $\delta \in [1.01,3.99]$, $R_{max} \in [\tilde{R},0.61]$, where $\tilde{R}=0.256$. Step size of grid = 0.01 in both the horizontal and vertical directions.}}
	\label{fig:mat12}
\end{figure}

\begin{figure}
	\centering
	\includegraphics[scale=0.65]{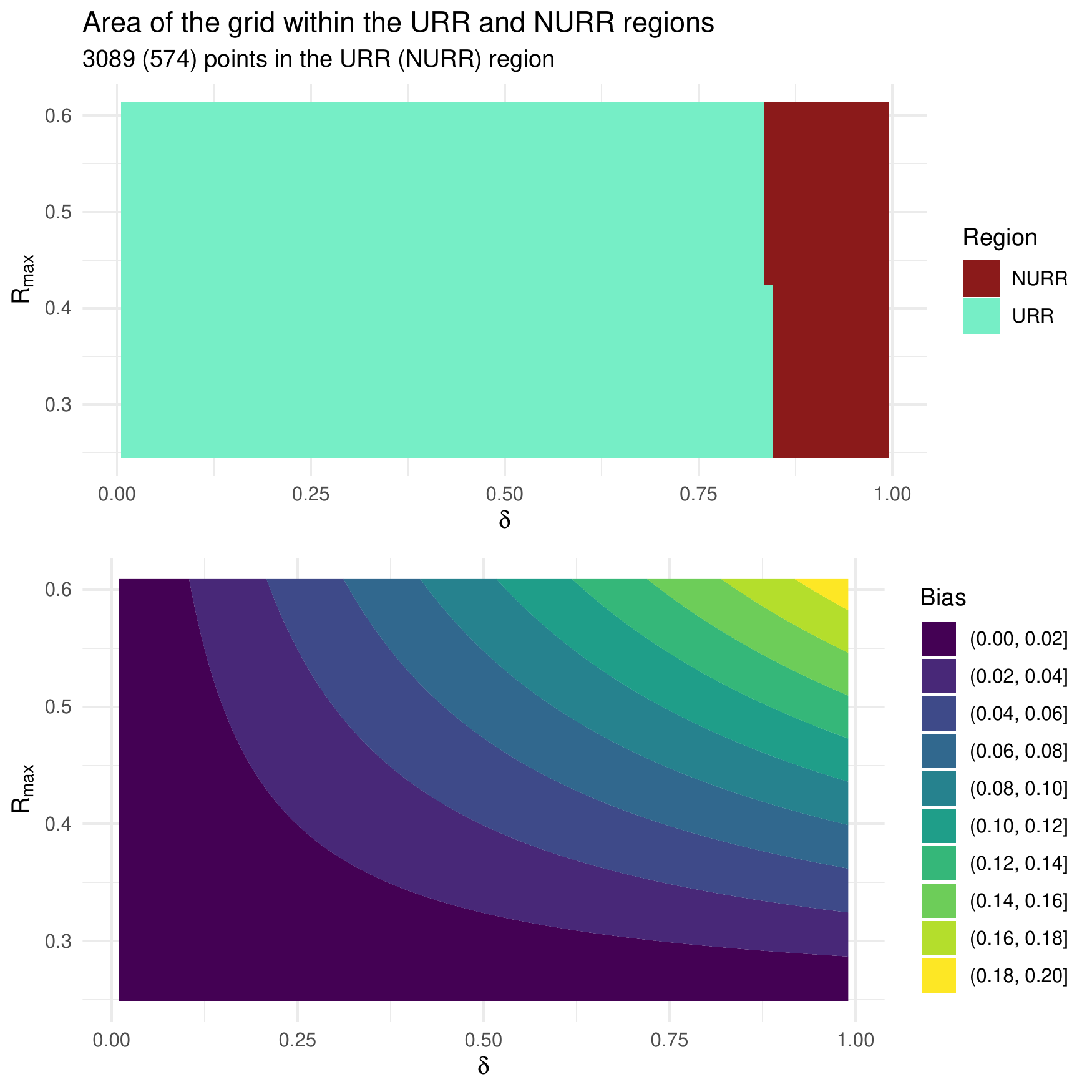}
	\caption{\textit{The figure identifies the region of unique real root (URR, top panel) and plots the contour of omitted variable bias (bottom panel) for the regression reported in row~2, Table~\ref{table:bset-maternal}. The bounded box is defined by $\delta \in [0.01,0.99]$, $R_{max} \in [\tilde{R},0.61]$, where $\tilde{R}=0.249$. Step size of grid = 0.01 in both the horizontal and vertical directions.}}
	\label{fig:mat21}
\end{figure}

\begin{figure}
	\centering
	\includegraphics[scale=0.65]{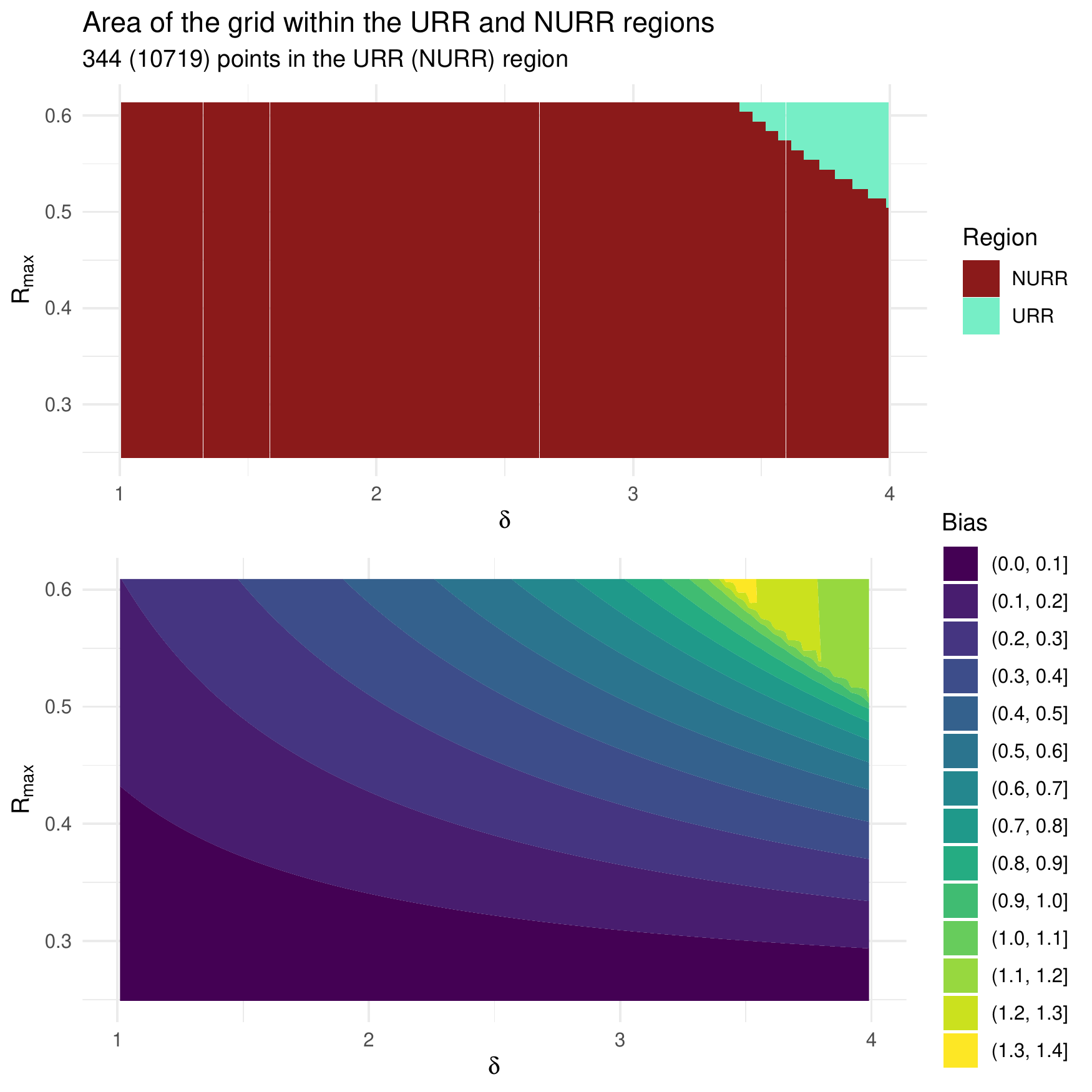}
	\caption{\textit{The figure identifies the region of unique real root (URR, top panel) and plots the contour of omitted variable bias (bottom panel) for the regression reported in row~2, Table~\ref{table:bset-maternal}. The bounded box is defined by $\delta \in [1.01,3.99]$, $R_{max} \in [\tilde{R},0.61]$, where $\tilde{R}=0.249$. Step size of grid = 0.01 in both the horizontal and vertical directions.}}
	\label{fig:mat22}
\end{figure}

\begin{figure}
	\centering
	\includegraphics[scale=0.65]{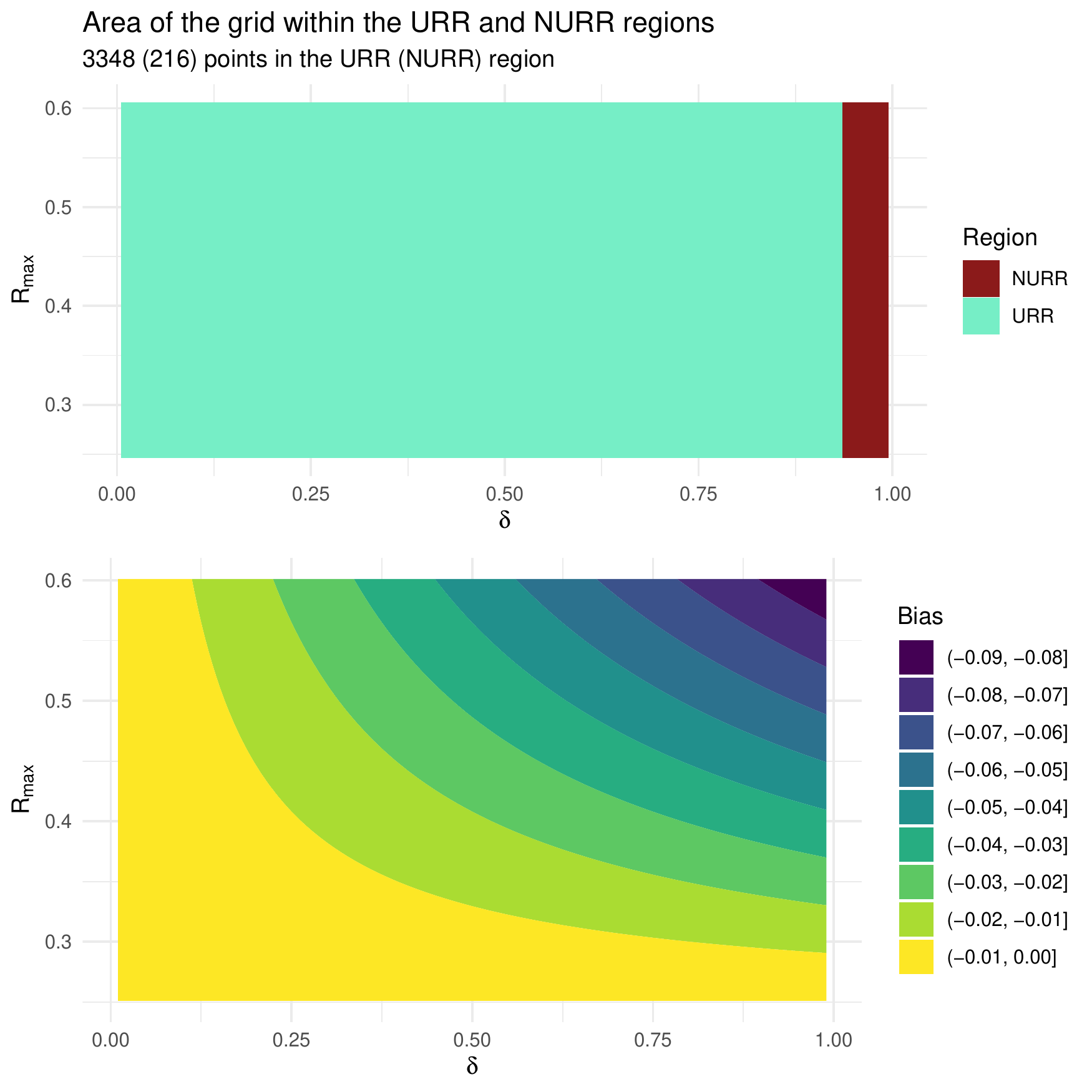}
	\caption{\textit{The figure identifies the region of unique real root (URR, top panel) and plots the contour of omitted variable bias (bottom panel) for the regression reported in row~3, Table~\ref{table:bset-maternal}. The bounded box is defined by $\delta \in [0.01,0.99]$, $R_{max} \in [\tilde{R},0.61]$, where $\tilde{R}=0.251$. Step size of grid = 0.01 in both the horizontal and vertical directions.}}
	\label{fig:mat31}
\end{figure}

\begin{figure}
	\centering
	\includegraphics[scale=0.65]{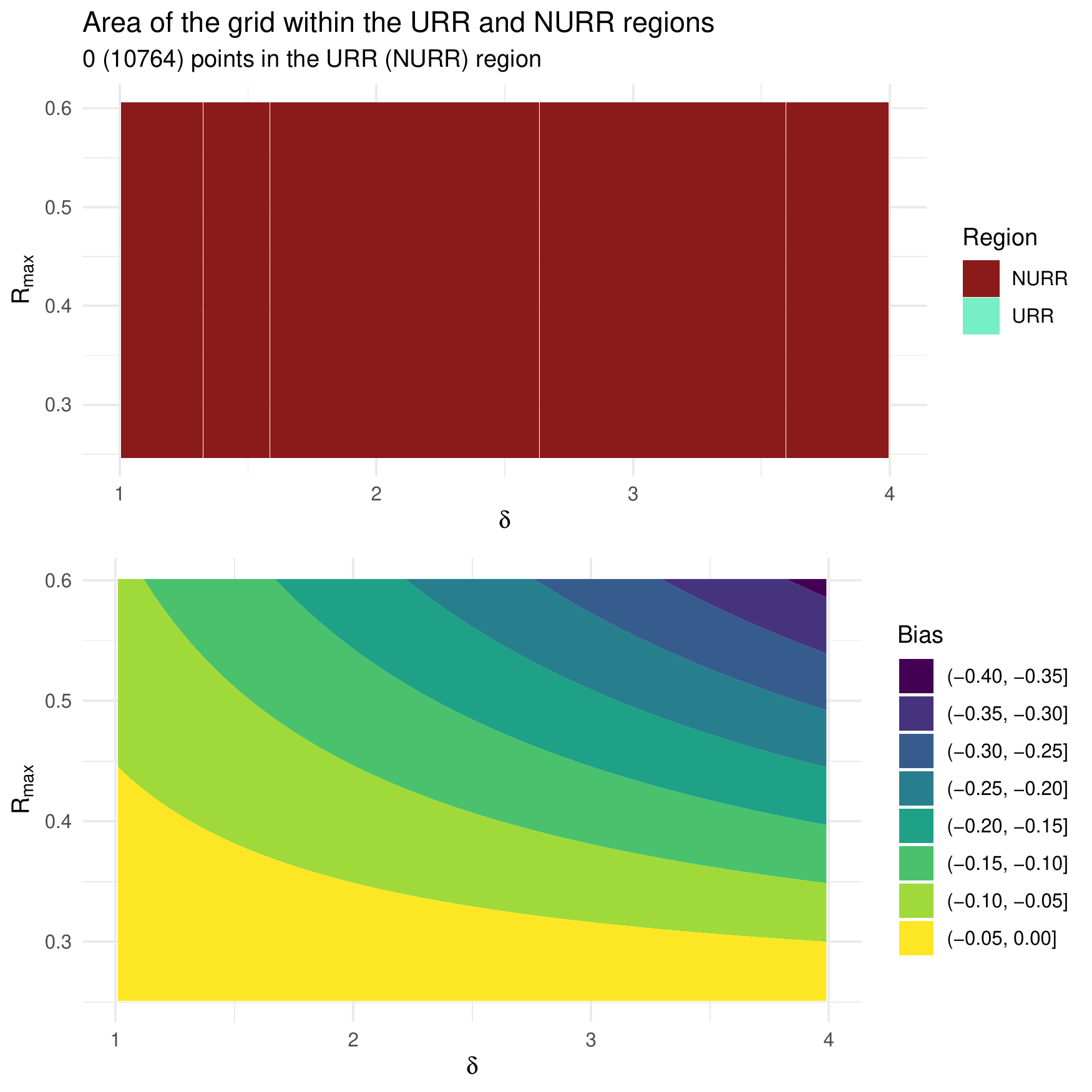}
	\caption{\textit{The figure identifies the region of unique real root (URR, top panel) and plots the contour of omitted variable bias (bottom panel) for the regression reported in row~3, Table~\ref{table:bset-maternal}. The bounded box is defined by $\delta \in [1.01,3.99]$, $R_{max} \in [\tilde{R},0.61]$, where $\tilde{R}=0.251$. Step size of grid = 0.01 in both the horizontal and vertical directions.}}
	\label{fig:mat32}
\end{figure}

\begin{figure}
	\centering
	\includegraphics[scale=0.65]{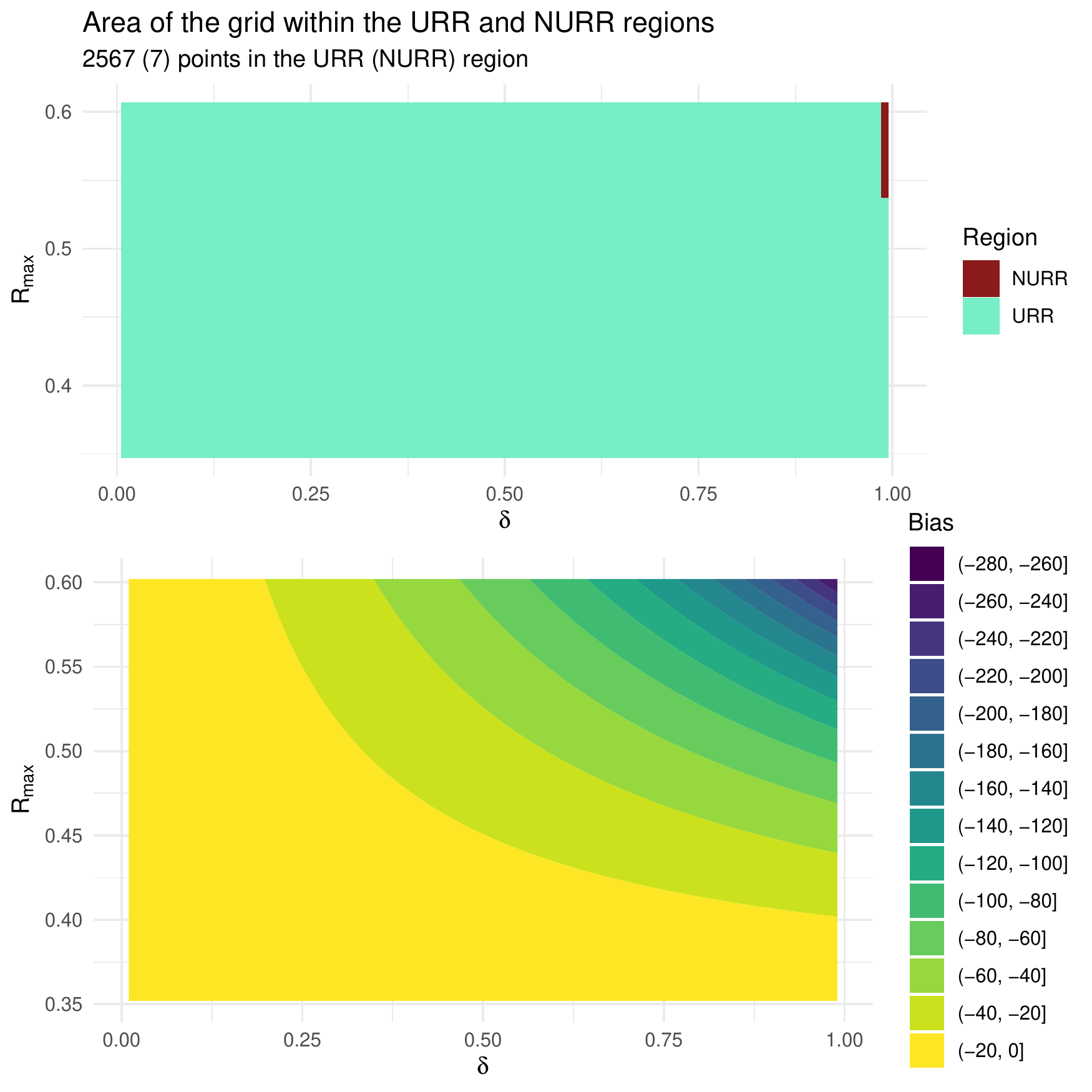}
	\caption{\textit{The figure identifies the region of unique real root (URR, top panel) and plots the contour of omitted variable bias (bottom panel) for the regression reported in row~4, Table~\ref{table:bset-maternal}. The bounded box is defined by $\delta \in [0.01,0.99]$, $R_{max} \in [\tilde{R},0.53]$, where $\tilde{R}=0.352$. Step size of grid = 0.01 in both the horizontal and vertical directions.}}
	\label{fig:mat41}
\end{figure}

\begin{figure}
	\centering
	\includegraphics[scale=0.65]{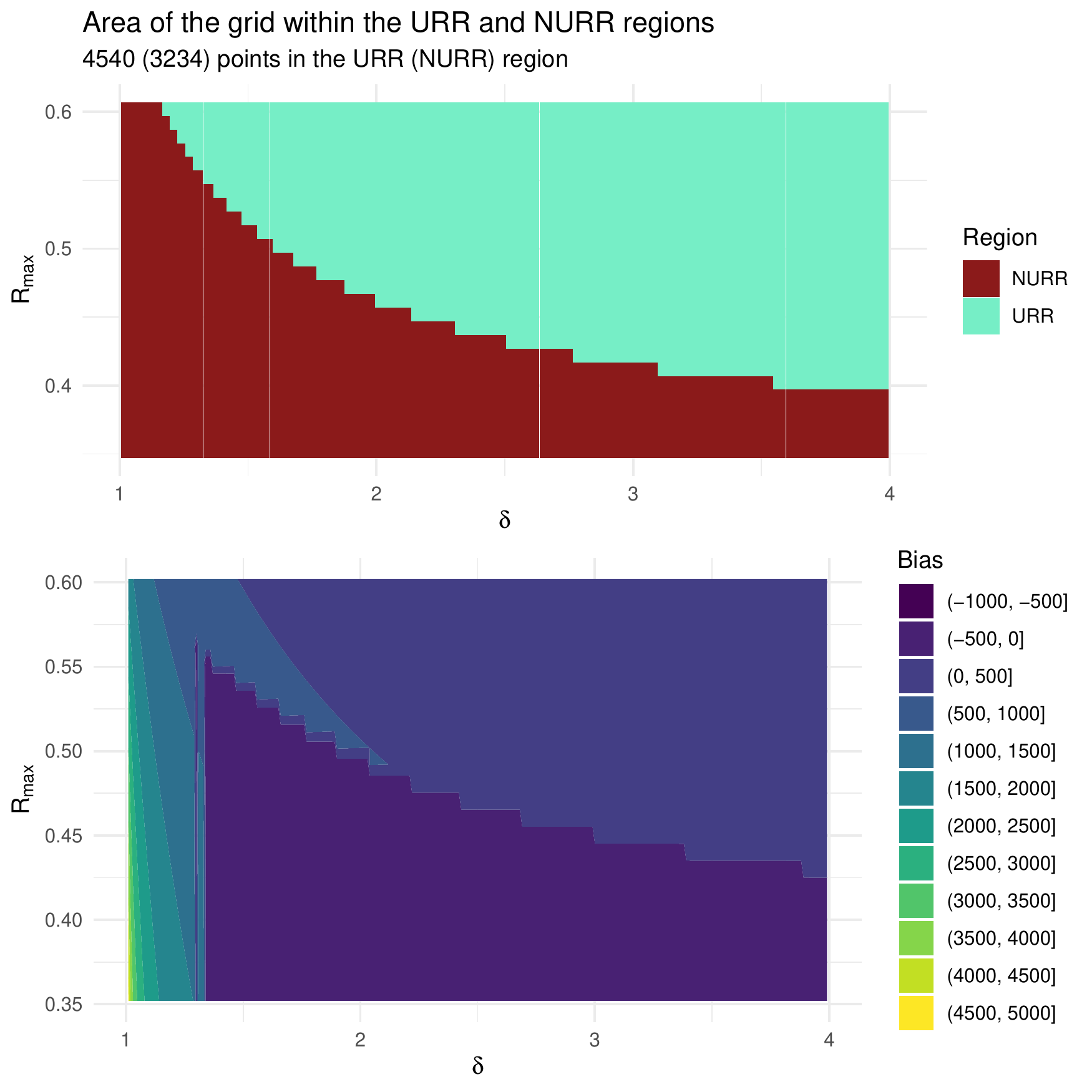}
	\caption{\textit{The figure identifies the region of unique real root (URR, top panel) and plots the contour of omitted variable bias (bottom panel) for the regression reported in row~4, Table~\ref{table:bset-maternal}. The bounded box is defined by $\delta \in [1.01,3.99]$, $R_{max} \in [\tilde{R},0.53]$, where $\tilde{R}=0.352$. Step size of grid = 0.01 in both the horizontal and vertical directions.}}
	\label{fig:mat42}
\end{figure}

\begin{figure}
	\centering
	\includegraphics[scale=0.65]{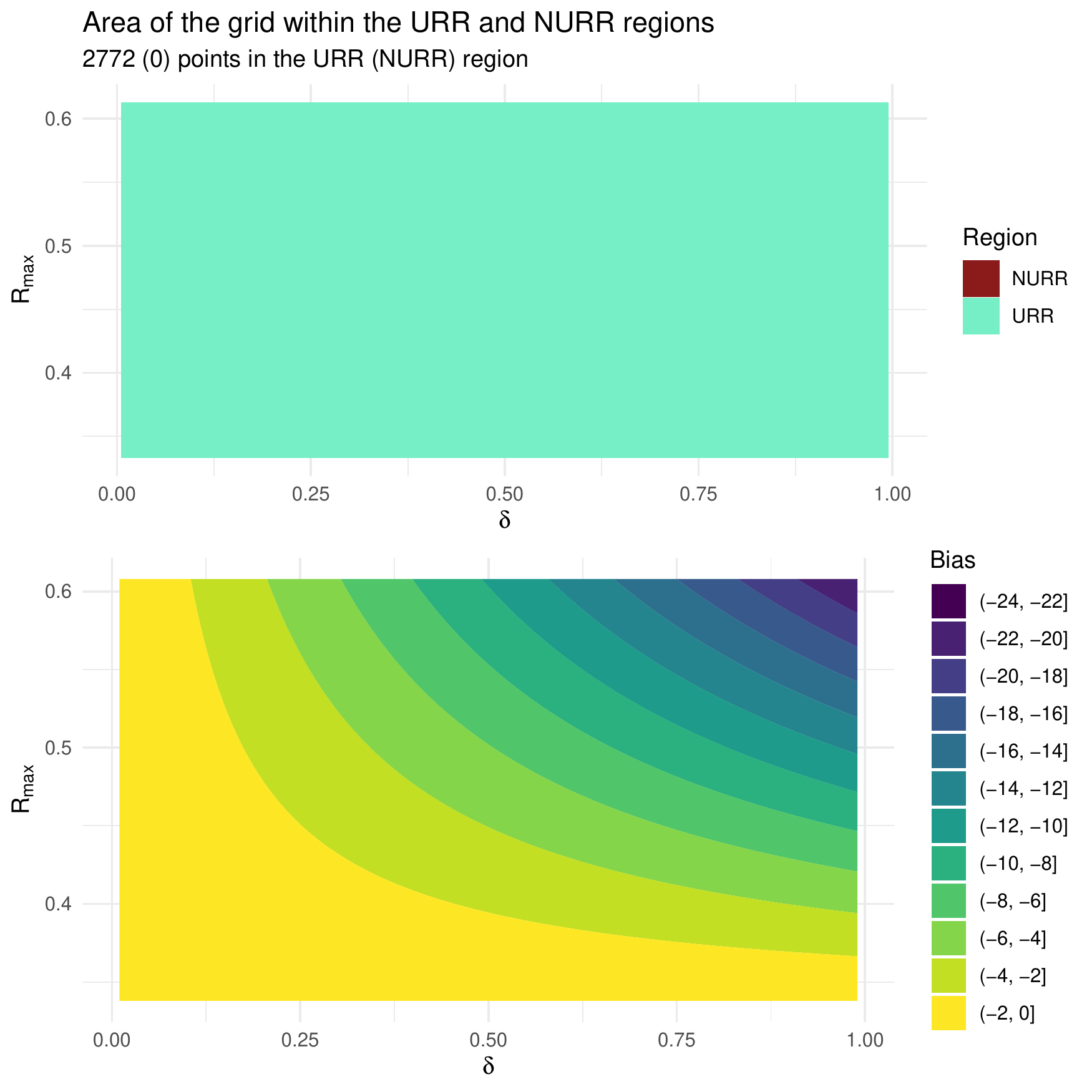}
	\caption{\textit{The figure identifies the region of unique real root (URR, top panel) and plots the contour of omitted variable bias (bottom panel) for the regression reported in row~5, Table~\ref{table:bset-maternal}. The bounded box is defined by $\delta \in [0.01,0.99]$, $R_{max} \in [\tilde{R},0.53]$, where $\tilde{R}=0.338$. Step size of grid = 0.01 in both the horizontal and vertical directions.}}
	\label{fig:mat51}
\end{figure}

\begin{figure}
	\centering
	\includegraphics[scale=0.65]{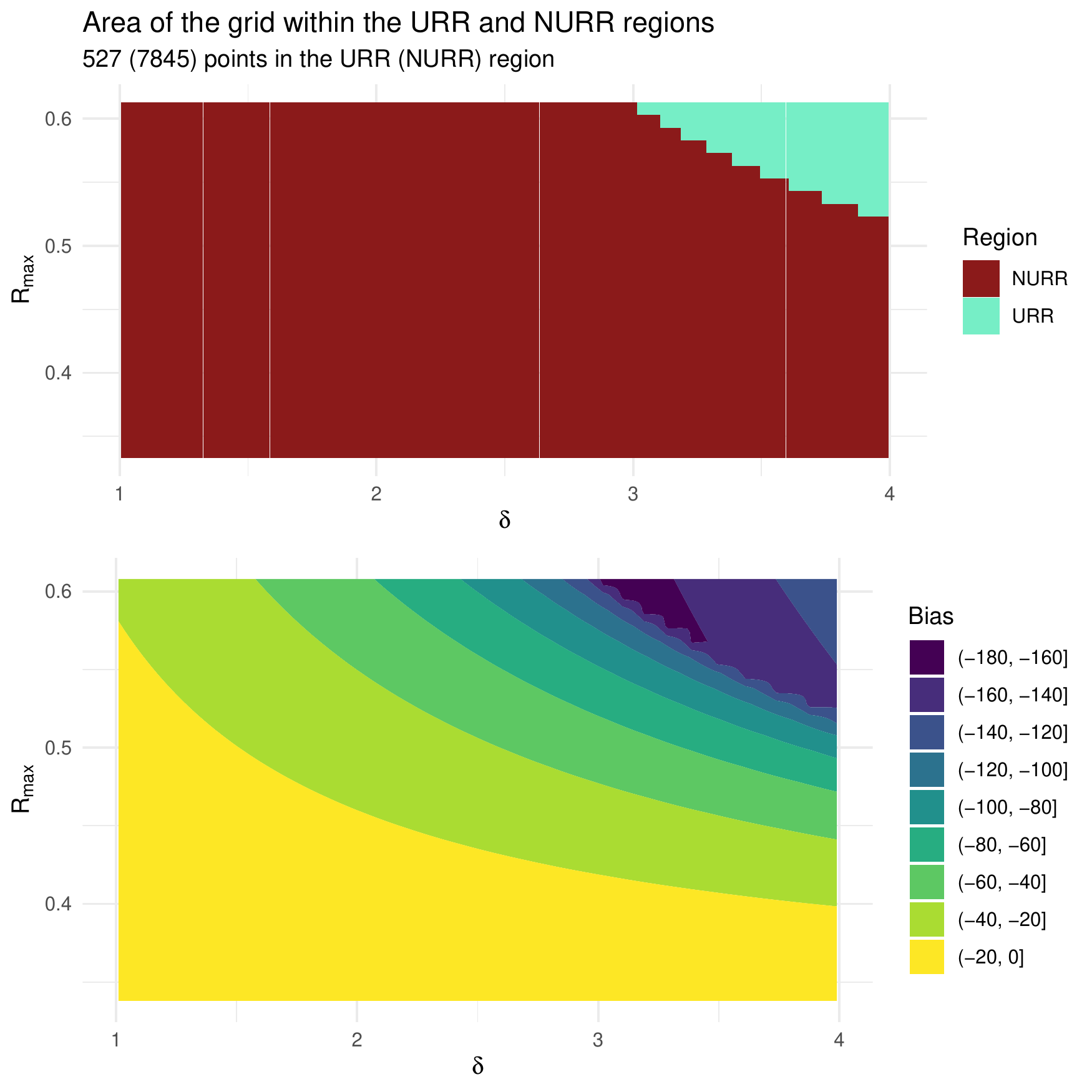}
	\caption{\textit{The figure identifies the region of unique real root (URR, top panel) and plots the contour of omitted variable bias (bottom panel) for the regression reported in row~5, Table~\ref{table:bset-maternal}. The bounded box is defined by $\delta \in [1.01,3.99]$, $R_{max} \in [\tilde{R},0.53]$, where $\tilde{R}=0.338$. Step size of grid = 0.01 in both the horizontal and vertical directions.}}
	\label{fig:mat52}
\end{figure}

\end{document}